\documentclass[acmsmall,screen,nonacm]{acmart}

\makeatletter
\newcommand{\TUItopfloats}{%
  \setcounter{topnumber}{1}%
  \setcounter{dbltopnumber}{1}%
  \setcounter{bottomnumber}{0}%
  \setcounter{totalnumber}{1}%
  \renewcommand{\topfraction}{0.9}%
  \renewcommand{\dbltopfraction}{0.9}%
  \renewcommand{\bottomfraction}{0}%
  \renewcommand{\textfraction}{0.1}%
  \renewcommand{\floatpagefraction}{0.5}%
  \setlength{\textfloatsep}{10pt plus 2pt minus 2pt}%
  \floatplacement{figure}{t}%
  \floatplacement{table}{t}%
  \setlength{\@fpsep}{\textheight}%
  \setlength{\@dblfpsep}{\textheight}%
}
\makeatother
\AtBeginDocument{%
  \TUItopfloats
}
\AddToHook{begindocument/end}{\TUItopfloats}

\usepackage{algorithmic}
\usepackage{colortbl}  
\usepackage{multirow}  
\usepackage{placeins}  
\usepackage{flafter}   
\usepackage{tabularx}  
\usepackage{wrapfig}   
\usepackage{subcaption} 
\usepackage{xspace}
\usepackage{tikz}      

\newcommand{\keycap}[1]{%
  \tikz[baseline=(key.base)]{
    \node[
      anchor=base,
      draw=black!35,
      fill=black!4,
      rounded corners=1pt,
      inner xsep=2.2pt,
      inner ysep=0.9pt,
      text height=1.25ex,
      text depth=0.25ex,
      line width=0.35pt,
      font=\footnotesize\ttfamily
    ] (key) {#1};
  }%
}

\newcommand{\AgentAsUser}{Agent-as-a-User\xspace}       
\newcommand{\FrameworkName}{\textsc{TUINaut}\xspace}       
\newcommand{\BenchmarkName}{\textsc{TUINaut-Bench}\xspace} 
\newcommand{\KeyboardAgent}{keyboard agent\xspace}     
\newcommand{\AgentKLM}{Agent-KLM\xspace}                
\newcommand{\InteractionController}{Interaction Controller\xspace} 
\newcommand{\InterpretationCost}{interpretation cost\xspace} 
\newcommand{\OperationCost}{operation cost\xspace}           
\newcommand{\InterpretationCostCap}{Interpretation cost\xspace} 
\newcommand{\ReasoningTokens}{reasoning tokens\xspace} 
\newcommand{\AtomicKeyEvents}{atomic key events\xspace}           

\setcopyright{none}
\begin{document}

\title{Beyond Task Completion: Measuring Interaction Cost in Terminal User Interfaces}

\author{Ruida Hu}
\email{200111107@stu.hit.edu.cn}
\affiliation{%
  \institution{Harbin Institute of Technology, Shenzhen}
  \city{Shenzhen}
  \country{China}}

\author{Yuanhao Wang}
\email{2023311I13@stu.hit.edu.cn}
\affiliation{%
  \institution{Harbin Institute of Technology, Shenzhen}
  \city{Shenzhen}
  \country{China}}

\author{Chao Peng}
\email{chao.peng@acm.org}
\affiliation{%
  \institution{University of Edinburgh}
  \city{Edinburgh}
  \country{United Kingdom}}

\author{Yakun Zhang}
\email{zhangyk@hit.edu.cn}
\affiliation{%
  \institution{Harbin Institute of Technology, Shenzhen}
  \city{Shenzhen}
  \country{China}}

\author{Cuiyun Gao}
\email{gaocuiyun@hit.edu.cn}
\affiliation{%
  \institution{Harbin Institute of Technology, Shenzhen}
  \city{Shenzhen}
  \country{China}}

\authorsaddresses{}

\providecommand{\RQOneHumanN}{6}
\providecommand{\RQOneHumanTaskRuns}{504}
\providecommand{\RQOneHumanFirstPasses}{385}
\providecommand{\RQOneHumanFirstPass}{76.4\%}
\providecommand{\RQOneAgentAttempts}{252}
\providecommand{\RQOneAgentSuccesses}{209}
\providecommand{\RQOneAgentPass}{82.9\%}
\providecommand{\RQOneAgentHelp}{13.5\%}
\providecommand{\RQOneHumanHelp}{14.7\%}
\providecommand{\RQOneSuccessRho}{0.35}
\providecommand{\RQOneSuccessRhoLow}{-0.03}
\providecommand{\RQOneSuccessRhoHigh}{0.57}
\providecommand{\RQOneCostRho}{0.77}
\providecommand{\RQOneCostRhoLow}{0.46}
\providecommand{\RQOneCostRhoHigh}{0.88}
\providecommand{\RQOnePrintableRho}{0.89}
\providecommand{\RQOnePrintableRhoLow}{0.65}
\providecommand{\RQOnePrintableRhoHigh}{0.94}
\providecommand{\RQOneNontextRho}{0.48}
\providecommand{\RQOneNontextRhoLow}{0.06}
\providecommand{\RQOneNontextRhoHigh}{0.71}
\providecommand{\RQOneHumanPrintableShare}{42.6\%}
\providecommand{\RQOneAgentPrintableShare}{73.6\%}
\providecommand{\RQOneHumanNavigationShare}{42.4\%}
\providecommand{\RQOneAgentNavigationShare}{14.2\%}
\providecommand{\RQOneHumanDeletionShare}{8.2\%}
\providecommand{\RQOneAgentDeletionShare}{0.9\%}
\providecommand{\RQOneHumanControlShare}{6.9\%}
\providecommand{\RQOneAgentControlShare}{11.3\%}
\providecommand{\RQOneReasoningRho}{0.62}
\providecommand{\RQOneReasoningRhoLow}{0.20}
\providecommand{\RQOneReasoningRhoHigh}{0.75}
\providecommand{\RQOneHumanAtomicMedian}{24.8}
\providecommand{\RQOneAgentAtomicMedian}{15.0}
\providecommand{\RQOneHumanPauseMedian}{46.0}
\providecommand{\RQOneAgentReasoningMedian}{3788}
\providecommand{\RQOneInterpretationTasks}{84}
\providecommand{\RQOneComparisonTasks}{84}
\providecommand{\RQOneHumanHumanSimilarity}{0.50}
\providecommand{\RQOneAgentAgentSimilarity}{0.92}
\providecommand{\RQOneHumanAgentSimilarity}{0.44}
\providecommand{\RQOneBothEasier}{60}
\providecommand{\RQOneBothHarder}{4}
\providecommand{\RQOneAgentFavored}{5}
\providecommand{\RQOneHumanFavored}{6}
\providecommand{\RQTwoSemanticDelta}{+1.2}
\providecommand{\RQTwoSemanticLow}{-4.12}
\providecommand{\RQTwoSemanticHigh}{+6.91}
\providecommand{\RQTwoPromptOverhead}{2.88}
\providecommand{\RQTwoKendallWSuccess}{0.7606}
\providecommand{\RQTwoKendallWSuccessLow}{0.5366}
\providecommand{\RQTwoKendallWSuccessHigh}{0.8566}
\providecommand{\RQTwoKendallWInterpretation}{0.8575}
\providecommand{\RQTwoKendallWInterpretationLow}{0.7544}
\providecommand{\RQTwoKendallWInterpretationHigh}{0.9080}
\providecommand{\RQTwoKendallWOperation}{0.8869}
\providecommand{\RQTwoKendallWOperationLow}{0.7599}
\providecommand{\RQTwoKendallWOperationHigh}{0.9597}
\providecommand{\RQTwoKendallWSuccessBody}{0.76}
\providecommand{\RQTwoKendallWSuccessLowBody}{0.54}
\providecommand{\RQTwoKendallWSuccessHighBody}{0.86}
\providecommand{\RQTwoKendallWInterpretationBody}{0.86}
\providecommand{\RQTwoKendallWInterpretationLowBody}{0.75}
\providecommand{\RQTwoKendallWInterpretationHighBody}{0.91}
\providecommand{\RQTwoKendallWOperationBody}{0.89}
\providecommand{\RQTwoKendallWOperationLowBody}{0.76}
\providecommand{\RQTwoKendallWOperationHighBody}{0.96}
\providecommand{\RQTwoKeyRatio}{1.05}
\providecommand{\RQTwoKeyRatioLow}{0.95}
\providecommand{\RQTwoKeyRatioHigh}{1.18}
\providecommand{\RQTwoReasoningRatio}{1.13}
\providecommand{\RQTwoReasoningRatioLow}{0.93}
\providecommand{\RQTwoReasoningRatioHigh}{1.39}
\providecommand{\RQTwoTurnRatio}{1.05}
\providecommand{\RQTwoTurnRatioLow}{0.96}
\providecommand{\RQTwoTurnRatioHigh}{1.16}
\providecommand{\RQTwoImageDelta}{-15.9}
\providecommand{\RQTwoImageLow}{-28.1}
\providecommand{\RQTwoImageHigh}{-4.4}
\providecommand{\RQTwoImageKeyRatio}{1.28}
\providecommand{\RQTwoImageKeyRatioLow}{1.04}
\providecommand{\RQTwoImageKeyRatioHigh}{1.54}
\providecommand{\RQTwoImageTurnRatio}{1.28}
\providecommand{\RQTwoImageTurnRatioLow}{1.08}
\providecommand{\RQTwoImageTurnRatioHigh}{1.51}
\providecommand{\RQTwoFlashSemanticDelta}{+0.8}
\providecommand{\RQTwoFlashSemanticLow}{-3.38}
\providecommand{\RQTwoFlashSemanticHigh}{+5.36}
\providecommand{\RQTwoFlashKeyRatio}{0.91}
\providecommand{\RQTwoFlashKeyRatioLow}{0.79}
\providecommand{\RQTwoFlashKeyRatioHigh}{1.03}
\providecommand{\RQThreeAllGeneratedFavored}{11}
\providecommand{\RQThreeReferenceFavored}{24}
\providecommand{\RQThreeReferencePass}{84.1\%}
\providecommand{\RQThreeReferenceRecovery}{67\%}
\providecommand{\RQThreeDeepSeekRecovery}{49\%}
\providecommand{\RQThreeClaudeCodeRecovery}{43\%}
\providecommand{\RQThreeCodexRecovery}{53\%}
\providecommand{\RQThreeDeepSeekPass}{75.4\%}
\providecommand{\RQThreeDeepSeekDelta}{-8.7}
\providecommand{\RQThreeDeepSeekLow}{-21.8}
\providecommand{\RQThreeDeepSeekHigh}{+3.6}
\providecommand{\RQThreeDeepSeekKeyRatio}{0.86}
\providecommand{\RQThreeDeepSeekKeyLow}{0.64}
\providecommand{\RQThreeDeepSeekKeyHigh}{1.17}
\providecommand{\RQThreeDeepSeekReasoningRatio}{0.58}
\providecommand{\RQThreeDeepSeekReasoningLow}{0.27}
\providecommand{\RQThreeDeepSeekReasoningHigh}{1.24}
\providecommand{\RQThreeDeepSeekWrongSubmit}{38}
\providecommand{\RQThreeClaudeCodePass}{85.7\%}
\providecommand{\RQThreeClaudeCodeDelta}{+1.6}
\providecommand{\RQThreeClaudeCodeLow}{-6.6}
\providecommand{\RQThreeClaudeCodeHigh}{+10.3}
\providecommand{\RQThreeClaudeCodeKeyRatio}{1.07}
\providecommand{\RQThreeClaudeCodeKeyLow}{0.75}
\providecommand{\RQThreeClaudeCodeKeyHigh}{1.60}
\providecommand{\RQThreeClaudeCodeReasoningRatio}{0.67}
\providecommand{\RQThreeClaudeCodeReasoningLow}{0.36}
\providecommand{\RQThreeClaudeCodeReasoningHigh}{1.27}
\providecommand{\RQThreeClaudeCodeWrongSubmit}{25}
\providecommand{\RQThreeCodexPass}{72.2\%}
\providecommand{\RQThreeCodexDelta}{-11.9}
\providecommand{\RQThreeCodexLow}{-26.7}
\providecommand{\RQThreeCodexHigh}{+1.0}
\providecommand{\RQThreeCodexKeyRatio}{0.81}
\providecommand{\RQThreeCodexKeyLow}{0.60}
\providecommand{\RQThreeCodexKeyHigh}{1.04}
\providecommand{\RQThreeCodexReasoningRatio}{0.58}
\providecommand{\RQThreeCodexReasoningLow}{0.37}
\providecommand{\RQThreeCodexReasoningHigh}{0.89}
\providecommand{\RQThreeCodexWrongSubmit}{43}
\providecommand{\RQThreeCompletionRhoMin}{0.22}
\providecommand{\RQThreeCompletionRhoMax}{0.44}
\providecommand{\RQThreeCompletionW}{0.59}
\providecommand{\RQThreeCompletionWLow}{0.42}
\providecommand{\RQThreeCompletionWHigh}{0.75}
\providecommand{\RQThreeInterpretationRhoMin}{0.29}
\providecommand{\RQThreeInterpretationRhoMax}{0.47}
\providecommand{\RQThreeInterpretationW}{0.62}
\providecommand{\RQThreeInterpretationWLow}{0.43}
\providecommand{\RQThreeInterpretationWHigh}{0.74}
\providecommand{\RQThreeOperationRhoMin}{0.76}
\providecommand{\RQThreeOperationRhoMax}{0.79}
\providecommand{\RQThreeOperationW}{0.84}
\providecommand{\RQThreeOperationWLow}{0.68}
\providecommand{\RQThreeOperationWHigh}{0.93}

\begin{abstract}
    With rapid advances in large language models (LLMs), terminal user interfaces (TUIs) have become increasingly popular for development and operations. 
    Successfully completing a task through a TUI does not mean that the task is easy to perform. 
    Because task success alone does not fully capture TUI usability, evaluation should also measure the effort involved in understanding and operating the interface.
    Despite this growing adoption, research on TUIs remains limited, and no prior work has systematically evaluated their usability.
    Existing usability assessments rely mainly on costly human assessment or LLM-generated ratings and reports. Neither provides a repeatable measure of task-level interaction effort grounded in actual task execution.

    
    To address these gaps, we propose \textbf{\AgentAsUser}, an evaluation paradigm that places an LLM agent in the user role to 
    provide repeatable measurements of task-level interaction effort.
    Specifically, we introduce \textbf{\AgentKLM}, an analytical model 
    that separates agent-side costs into interpretation and operation dimensions to establish a structural correspondence with human interaction.
    To operationalize the \AgentAsUser paradigm,
    we implement \textbf{\FrameworkName}, a framework that records realized interaction trajectories and verifies task outcomes using actor-independent oracles.
    
    We construct \textbf{\BenchmarkName} and conduct an empirical study of 84 tasks across 15 real-world TUIs.
    The study includes six human participants, five observation--evaluator configurations, 
    and we apply the paradigm to 45 TUIs independently generated by three leading generation stacks.
    We find that similar aggregate success rates can mask substantial differences in which tasks humans and agents complete, whereas their interpretation and operation costs follow correlated task rankings, especially for operation.
    Across all five observation--evaluator configurations, completion correspondence remains weak while cost rankings remain aligned.
    TUIs generated by current leading agent frameworks can implement correct functionality, yet their task-based usability still needs improvement.
    These findings position \AgentAsUser as a repeatable and execution-grounded paradigm for measuring task-based usability.

    All artifacts for \FrameworkName and \BenchmarkName are available at \url{https://github.com/kinesiatricssxilm14/Agent-as-a-User}. Anonymized trajectories for all runs are available at \url{https://tuinaut.pages.dev/}.
\end{abstract}


\maketitle
\TUItopfloats

\section{Introduction}
The spread of large language models (LLMs) has given terminal user interfaces (TUIs) a larger role in development and operations. 
These interfaces now take many forms, such as file managers, system monitors, database clients, version-control tools, and coding agents~\cite{DBLP:journals/corr/abs-2608-03743,DBLP:conf/iclr/MerrillSCLRBSSW26,DBLP:journals/corr/abs-2606-28480,rocha2023charm,anthropic2025claudecode}. 
Compared with conventional graphical user interfaces (GUIs), TUIs
render their interfaces in terminals and rely more heavily on keyboard input,
often through mode-dependent keys with limited visible cues.
These characteristics make it important to study how easily users can complete concrete tasks.
However, research on TUI usability remains limited.

\begin{figure*}[t]
    \centering
    \includegraphics[width=\textwidth]{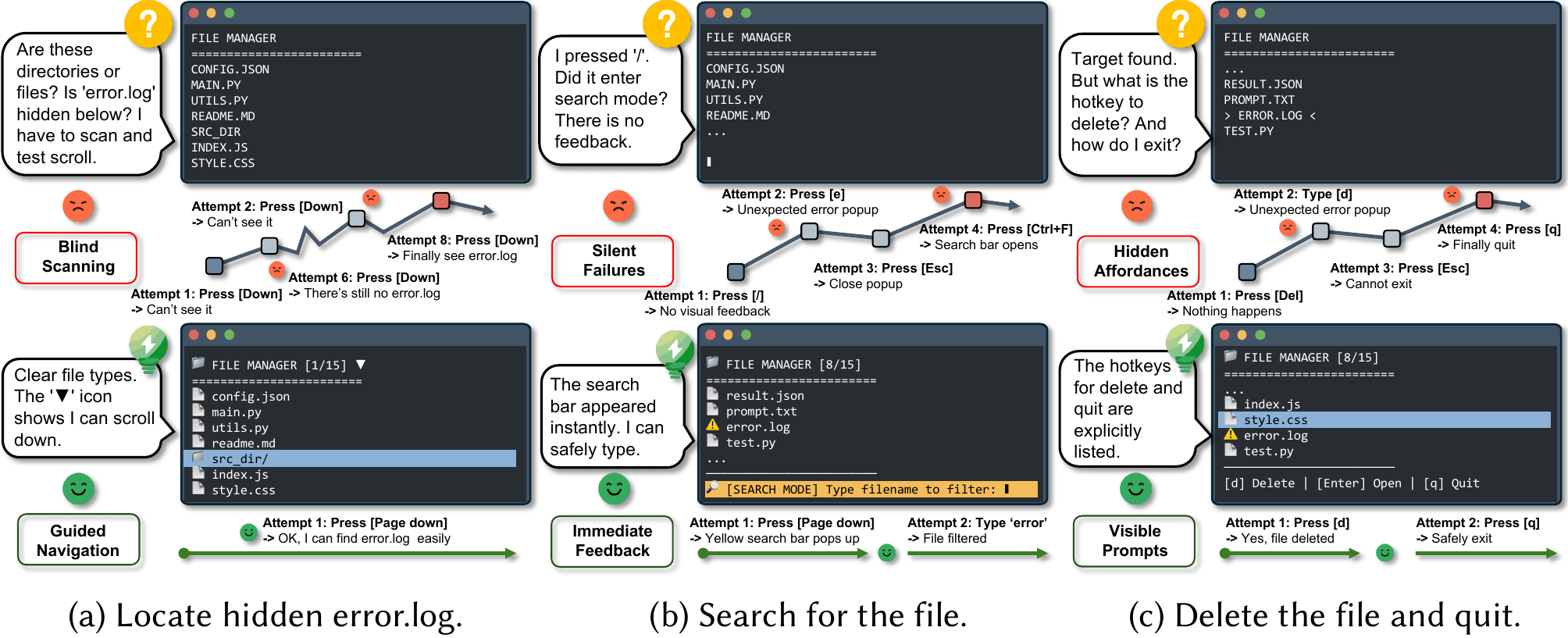}
    \caption{The same file-management goals can expose different interaction costs. Hidden scroll, search-state, and command cues produce blind scanning, trial-and-error input, and silent failures (top). Guided navigation, immediate feedback, and visible prompts support direct action (bottom).}
    \Description{Two rows compare three file-manager goals: locating a hidden file, searching, and deleting the file before quitting. The upper interfaces omit scroll, search-state, and command cues, causing repeated unsuccessful actions. The lower interfaces expose these cues and enable shorter, guided paths.}
    \label{fig:intro_motivation}
\end{figure*}

How easily users can complete a task depends on the cues exposed by the interface. Fig.~\ref{fig:intro_motivation} compares two file-manager designs for the same goals.
In the lower design, guided navigation in Fig.~\ref{fig:intro_motivation}(a) reveals that more entries lie below and indicates where to scroll. Immediate feedback in Fig.~\ref{fig:intro_motivation}(b) confirms that the search has started. Visible prompts in Fig.~\ref{fig:intro_motivation}(c) expose the keys for delete and quit.
The upper design omits these cues, forcing users to scan blindly, try actions, and infer whether those actions succeeded.
The same tasks can therefore impose different demands on understanding the interface and acting through it.

These examples show that task completion alone cannot characterize \textbf{task-based usability}. We therefore consider two complementary forms of evidence from actual executions: task completion, which records whether the goal is reached, and interaction cost, which captures the demands of interpreting the interface state and carrying out the next action.

Meanwhile, software evaluation still focuses primarily on correctness~\cite{DBLP:conf/iclr/JimenezYWYPPN24,DBLP:journals/corr/abs-2509-16941,DBLP:conf/nips/ZanHLCXZLLCZLXC25,DBLP:conf/iclr/JainHGLYZWSSS25}, and usability assessments leave the interaction cost of a task unmeasured. 
Existing usability assessments rely mainly on human studies or LLM-generated ratings and reports.
Human studies can reveal interaction friction and subjective experience, but they are expensive to repeat across applications and revisions and difficult to scale~\cite{DBLP:journals/csur/IvoryH01,DBLP:conf/chi/NielsenL93,DBLP:conf/chi/JeffriesMWU91,DBLP:conf/chi/KuangJ022}.
Recent LLM-based systems simulate users, generate think-aloud reports, predict ratings, or organize UX feedback~\cite{DBLP:conf/chi/0003YGHWLG0LW25,DBLP:conf/hicss/GriesbachSWG26,DBLP:journals/corr/abs-2601-15777}. 
These systems provide useful qualitative evidence, but their ratings and reports remain judgments rather than measurements grounded in verified task execution.
Therefore, existing approaches do not provide a repeatable measure of interaction cost grounded in actual task execution.

To address this measurement problem, we propose \textbf{\AgentAsUser}, an evaluation paradigm that places an LLM agent in the user role to evaluate task-based TUI usability. 
The agent executes each task under a fixed configuration, with a fresh context and keyboard actions only.
Under this setup, the agent serves as a controlled experimental instrument that observes the same rendered TUI and acts through the same application input boundary as a human user.
Building on the classical Keystroke-Level Model (KLM), we introduce \textbf{\AgentKLM}, an analytical model that separates agent-side \textbf{\InterpretationCost} and \textbf{\OperationCost} and relates them structurally to human interaction. 
We implement \textbf{\FrameworkName}, a framework that records realized interaction trajectories and verifies task outcomes using actor-independent oracles.
Together, the paradigm, analytical model, and framework support an empirical study organized around three research questions.
\textbf{RQ1} asks whether agent completion and the two costs correspond to human interaction on the same tasks. \textbf{RQ2} asks whether that correspondence still holds for plain, semantic, and image observations, because humans and agents differ most in what they see. \textbf{RQ3} applies the paradigm to the task-based usability of TUIs generated by three leading coding agents.

To answer these questions, we construct \textbf{\BenchmarkName} with 84 tasks across 15 real-world TUIs.
The study includes six human participants, five observation--evaluator configurations spanning plain, semantic, and image input, 3,528 agent attempts, and 45 TUIs independently generated by the three stacks. 

The results show that similar aggregate success rates can hide substantial differences in which tasks humans and agents complete.
By contrast, their interpretation and operation costs follow correlated task rankings, with the strongest relationship for operation. 
This pattern recurs across the five configurations even though completion correspondence remains weak. 
Applying \AgentAsUser to generated TUIs further shows that correct functionality can coexist with higher interpretation cost and no consistent operation advantage. 
Together, these findings position \AgentAsUser as a repeatable and execution-grounded paradigm for measuring task-based usability.

This paper makes three contributions:
\begin{itemize}
    \item 
    \textbf{An evaluation paradigm for task-based TUI usability.}
    We propose \AgentAsUser and empirically characterize its human correspondence, stability across observation settings, and application to generated TUIs.
    \item 
    \textbf{An interaction-cost model for agent task execution.}
    We introduce \AgentKLM to separate interpretation and operation costs and operationalize it in \FrameworkName through verified task execution, alternative screen representations, atomic key-event accounting, repeated trials, and trajectory analysis.
    \item \textbf{A reproducible TUI benchmark.} We construct \BenchmarkName from 84 validated tasks over 15 recent real-world repositories, with reproducible fixtures, generation prompts, fingerprints, reference executions, and deterministic multi-level oracles.
\end{itemize}

\section{\AgentKLM: Modeling Interaction Cost for Task-Based Usability}
\label{sec:agent-profile}

\begin{figure}[t]
    \centering
    \includegraphics[width=\textwidth]{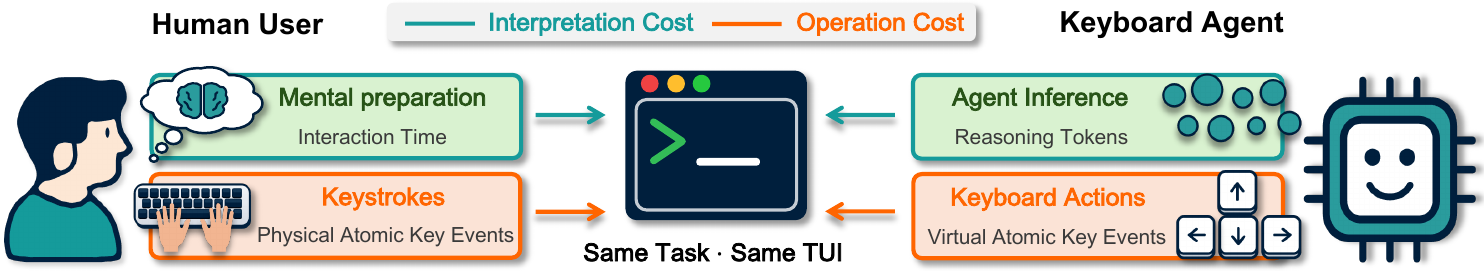}
    \caption{\InterpretationCostCap and \OperationCost for human and agent execution of the same TUI task. \AgentKLM defines the agent-side cost profile.}
    \Description{A human user and a keyboard agent face the same central terminal interface and task. Teal cards map human Interaction Time and agent Reasoning Tokens to Interpretation Cost. Orange cards map physical and virtual atomic key events to Operation Cost. Arrows connect all four mappings to the shared TUI.}
    \label{fig:agent-klm-overview}
\end{figure}

We introduce \textbf{\AgentKLM}, an analytical model that separates agent execution into \textbf{\InterpretationCost} and \textbf{\OperationCost}. This separation draws on classical KLM's distinction between mental preparation and physical operators. We first review this basis and then define the agent costs and corresponding human measures. Fig.~\ref{fig:agent-klm-overview} summarizes their relationship for the same TUI task.

\subsection{Classical KLM: Basis for Cost Decomposition}
Classical KLM starts from a simple observation: interfaces that support the same goal may still require different amounts of preparation and physical action~\cite{card1980keystroke}. It makes this difference explicit by decomposing a task method into elementary operators, where each operator denotes a primitive unit of user activity~\cite{card2018psychology}. The keystroke operator ($K$) denotes a key or button press. Pointing ($P$) moves a pointer to a target, homing ($H$) moves the hands between input devices, and drawing ($D$) traces a path. The mental-preparation operator ($M$) precedes an action or action chunk, while response ($R$) is system processing that blocks the next action. Assigning a duration to each operator yields an additive prediction of skilled human execution time:
\begin{equation}
T_{H} =
\underbrace{n_M t_M}_{\text{mental-preparation operator}}+
\underbrace{n_K t_K}_{\text{keystroke operator}}+
\underbrace{n_P t_P+n_H t_H+T_D}_{\text{pointing, homing, and drawing operators}}+
\underbrace{T_R}_{\text{blocking response}} .
\label{eq:classic-klm}
\end{equation}
Here, $n_X$ is the number of occurrences of operator $X$, $t_X$ is the duration of one such occurrence, $T_D$ is the total drawing duration, and $T_R$ is the total blocking system-response duration. Because the model predicts execution time, all $t$ and $T$ quantities are measured in seconds. The $M$ operator represents routine preparation before an action or action chunk. It is not intended to model unrestricted comprehension, learning, or problem solving~\cite{DBLP:journals/tochi/JohnK96a}.
\AgentKLM adopts this decomposition, using role-specific interpretation and operation measures rather than additive time prediction.

\subsection{\AgentKLM Interaction Cost}
\label{sec:agent-klm-cost}
The two dimensions provide a shared analytical structure for describing human and agent interaction. \InterpretationCostCap captures the resources used to determine what to do, whereas \OperationCost captures the keyboard input used to execute the chosen action. Classical KLM motivates this separation. At a keyboard-controlled interface,\footnote{Although some TUIs support mouse input, keyboard input is the more consistently available interaction mode. Restricting both humans and agents to it therefore enables consistent action measurement.} the pointing and drawing operators do not apply, and no homing is required between input devices. The response term $T_R$ captures system processing rather than actor-side activity, so it is not included in either dimension.

Fig.~\ref{fig:agent-klm-overview} maps the two dimensions to role-specific human and agent measures.
For humans, interaction time from the initial screen to the first submission provides an observable proxy for interpretation demand, while total physical atomic key events quantify operation. This interval does not isolate cognitive effort, but approximates the time required to understand, plan, and complete the interaction. For agents, reasoning tokens across all turns quantify inference, while total virtual atomic key events quantify operation. The human measures support correspondence analysis but are not components of \AgentKLM. \AgentKLM uses the two dimensions to define agent-side costs rather than predict execution time. 

We represent human or agent interaction over $m$ turns as the ordered trajectory
\begin{equation}
\tau = \bigl\langle(o_i,r_i,a_i)\bigr\rangle_{i=1}^{m},
\label{eq:interaction-trajectory}
\end{equation}
where, at turn $i$, $o_i$ is the screen observation, $r_i$ is the role-specific interpretation process, and $a_i$ is the emitted action. For the agent path, this separation mirrors ReAct~\cite{DBLP:conf/iclr/YaoZYDSN023}: reasoning updates the agent's plan, whereas actions change the external environment.

Within the agent trajectory, each interpretation process $r_i$ is recorded as a sequence of \textbf{reasoning tokens}. Its token count therefore provides an observable measure of inference expenditure. This choice is consistent with prior work, which uses reasoning length to characterize test-time computation and quantify inference cost~\cite{DBLP:conf/iclr/Snell0XK25,DBLP:conf/emnlp/MuennighoffYSLFHZLCH25,DBLP:conf/acl/HanWFZM025,DBLP:conf/ijcnlp/LeeRV25}. As an agent-native measure, reasoning-token count is compared with human interaction time under the shared \InterpretationCost dimension, but the two measures retain different meanings and units.

Within the same agent trajectory, each action $a_i$ is expanded into virtual \textbf{atomic key events}. Rather than treating the action as one high-level step, the resulting count records each individual key event delivered to the TUI and is reported in atomic key events (AKE). Physical atomic key events from human input follow the same counting rule. Using reasoning-token and atomic key event counts, we define the two agent-side costs as
\begin{align}
C^{\mathrm{int}}(\tau) &= \sum_{i=1}^{m} \lvert r_i\rvert_{\mathrm{tok}},
\label{eq:reasoning-expenditure}\\
C^{\mathrm{op}}(\tau) &= \sum_{i=1}^{m} \lvert a_i\rvert_{\mathrm{key}}.
\label{eq:agent-key-events}
\end{align}
Here, $C^{\mathrm{int}}$ is the \InterpretationCost and $C^{\mathrm{op}}$ is the \OperationCost. The terms $\lvert r_i\rvert_{\mathrm{tok}}$ and $\lvert a_i\rvert_{\mathrm{key}}$ count the reasoning tokens in $r_i$ and the virtual atomic key events represented by $a_i$, respectively.
Following classical KLM, both physical and virtual input are counted as individual key events rather than resulting characters: modifier keys such as \keycap{Shift} and \keycap{Ctrl} count as separate operations~\cite{card1980keystroke}. To make character-to-key expansion reproducible, we apply the widely adopted US QWERTY mapping to both human and agent input. The mapping follows the W3C \textit{UI Events KeyboardEvent code Values} specification, which identifies physical key positions independently of the resulting characters~\cite{w3c2025uieventscode}. A key press without a modifier contributes one event. When an input requires a modifier, the modifier and the main key are counted separately. For example, \keycap{?} is expanded into \keycap{Shift}+\keycap{/} and contributes two events. Agent batches are expanded in the same way. For example, \texttt{Abc} is counted as the virtual sequence \keycap{Shift}+\keycap{A}, \keycap{B}, and \keycap{C}, totaling four events. This makes $\lvert a_i\rvert_{\mathrm{key}}$ follow the same counting rule as physical human input.

Reasoning-token counts and operation counts reported in AKE use different units, so we do not add the two costs directly. Instead, \AgentKLM reports them as the ordered profile
\begin{equation}
\mathbf{C}(\tau) =
\left\langle C^{\mathrm{int}}(\tau), C^{\mathrm{op}}(\tau) \right\rangle .
\label{eq:agent-cost-profile}
\end{equation}
Here, $\mathbf{C}$ denotes the overall \AgentKLM cost profile. Its two components represent interface interpretation and keyboard operation, the two core aspects of agent interaction modeled by \AgentKLM. Keeping the two components separate allows us to distinguish whether an agent's interaction cost arises primarily from interpreting the interface or from operating it through keyboard input.

\section{\FrameworkName: A Framework for \AgentAsUser Evaluation}
\label{sec:method}

\begin{figure}[t]
    \centering
    \includegraphics[width=\linewidth]{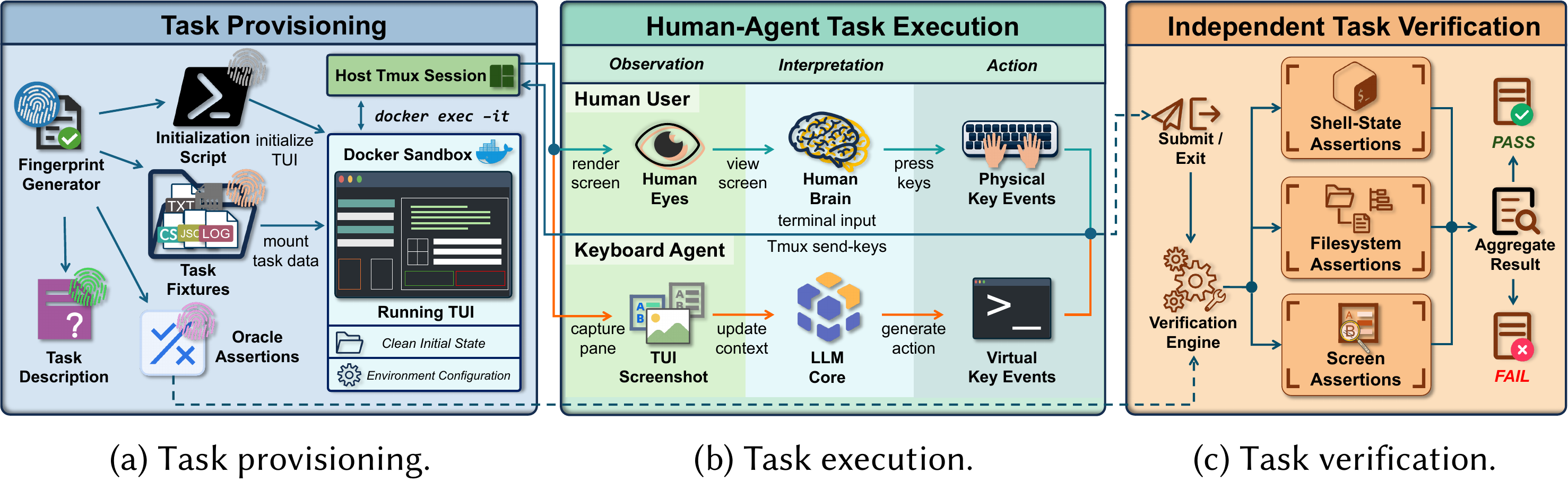}
    \caption{Overview of the \FrameworkName workflow. Human and agent users execute TUI tasks in aligned roles, while actor-independent oracles verify the resulting task states.}
    \Description{Task assets initialize a sandboxed TUI. A human user or keyboard agent observes its rendered state, interprets it through role-specific processes, and issues actions through keyboard events. Filesystem, shell-state, and screen oracles independently verify completion.}
    \label{fig:framework}
\end{figure}

To conduct \AgentAsUser evaluations and analyze the resulting trajectories with \AgentKLM, we develop the evaluation framework \FrameworkName and construct the task benchmark \BenchmarkName. 

\subsection{\BenchmarkName Construction}
\label{sec:benchmark-construction}
\subsubsection{Repository Selection}
\label{sec:repository-selection}
We begin with a broad GitHub search for TUI repositories created in 2026. This cutoff is intended to minimize potential overlap with model pretraining data while retaining a sufficiently large pool for screening. The queries use the \texttt{tui} and \texttt{terminal-ui} tags, supplemented by TUI framework names and application terms. The search returns 1,980 unique repositories.\footnote{We conducted the repository crawl in May 2026.} Requiring more than 50 stars leaves 274 candidates. Two authors then independently install and conduct hands-on evaluations of all 274 candidates, following their README workflows to check the claimed core functionality. This evaluation identifies candidates that launch reliably in Docker and whose main interactive features work as documented.

From this pool, we select 50 applications covering diverse domains and interaction patterns for in-depth examination using four criteria. \textbf{(1) Verifiability:} completion states are externally verifiable. \textbf{(2) Reproducibility:} repeating the same workflow from the same initial state produces consistent outcomes. \textbf{(3) Discoverability:} essential operations are discoverable within the TUI without external instructions. \textbf{(4) Diversity:} the selected set covers varied functionality and interaction patterns. When multiple repositories offer substantially similar functionality, we retain only one so that the final set includes a wider variety of TUIs. The two authors then compare and discuss their assessments in depth before jointly selecting 15 repositories. A third evaluator then independently verifies that each selected application launches and that its principal functions work as documented. Importantly, our final selection follows these four criteria and does not use subjective judgments of ease of use, which would narrow the range of TUIs represented in \BenchmarkName.

\subsubsection{Task Construction}
\label{sec:task-construction}
\begin{figure}[t]
    \centering
    \includegraphics[width=\textwidth]{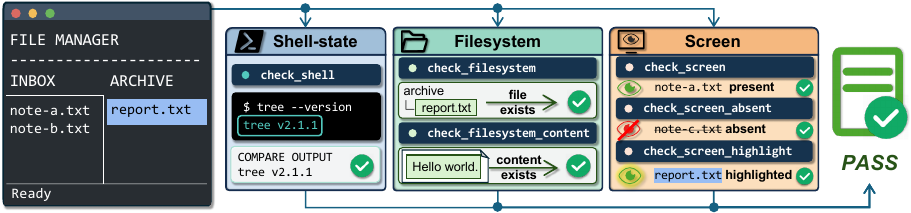}
    \caption{Six oracle assertion primitives in \BenchmarkName and the evidence inspected by each.}
    \Description{An example file-manager state connects to shell, filesystem, and screen assertion panels. The shell panel runs tree version and compares its output. The filesystem panel checks that archive report text exists and contains Hello world. The screen panel checks for present, absent, and highlighted text. Green check marks lead to a PASS result.}
    \label{fig:oracle-primitives}
\end{figure}

For reproducibility, we pin each repository's commit SHA and dependency versions. We also write a Dockerfile that installs its dependencies and provides a reproducible execution environment~\cite{DBLP:journals/corr/Boettiger14}. Over two months, the two authors carefully discuss and manually design all 84 tasks around each TUI's distinctive characteristics and principal functionality. Each task package includes an initialization script that establishes the starting state, a task description that specifies the user goal, task fixtures that supply the required data, and oracle assertions that verify completion, as shown in Fig.~\ref{fig:framework}(a). For each task, we complete the intended workflow by operating the TUI inside its Docker container and derive the oracle assertions from the expected final state. We manually repeat every task at least five times, at different times and on different machines, to confirm that it is reproducible and has a reachable successful path. We also verify that the information available within the TUI is sufficient to complete the task.


\subsubsection{Generation Prompt Construction}
\label{sec:generation-prompt-construction}
To support controlled generation of comparable TUIs, we construct one repository-level prompt for each selected application. Two authors manually derive each specification from the README, hands-on inspection, the functionality covered by its tasks, the technology stack, and the execution environment. Following recent machine-assisted benchmark construction and validation practices~\cite{DBLP:conf/nips/YangLJWKZHPSY25,DBLP:conf/nips/BadertdinovGNSK25,DBLP:conf/acl/PengSWZSG26,DBLP:conf/acl/ZhangGWZWLCSMZZ26}, an LLM helps organize this material under a shared template covering the project scenario, required capabilities, technical and environmental constraints, UI guidelines, and delivery rules. The two authors independently check each draft against the source materials and finalize it only after confirming that all required capabilities are clearly stated. The prompts specify only the required functionality and development constraints, without exposing benchmark tasks or the implementation and interaction design of the reference TUI. We release all 15 prompts as part of \BenchmarkName.

Overall, \BenchmarkName contains 84 tasks and 15 generation prompts across 15 repositories created between January and May 2026.\footnote{All benchmark data, including detailed statistics, are available in our repository and web page.}

\subsection{End-to-End Evaluation with \FrameworkName}
\BenchmarkName provides the applications and task assets, while \FrameworkName turns them into executable and verifiable evaluations. Fig.~\ref{fig:framework} summarizes this process in three stages.

\subsubsection{Task Provisioning}
\label{sec:task-provisioning}
As shown in Fig.~\ref{fig:framework}(a), \FrameworkName prepares each task in an isolated Docker environment and exposes its TUI through a host-managed Tmux session~\cite{marriott2026tmux}. It first starts the benchmark image as a long-lived Docker container and materializes the task fixtures at their designated paths in the Docker sandbox. \FrameworkName then runs \texttt{docker exec -it} from the Tmux session to start the task through a persistent interactive terminal shared by human and agent execution. Inside the container, the entrypoint runs the initialization script to reset task state, configure the environment, and prepare the fixture data before launching the TUI. Provisioning finishes when the initial screen is rendered. Every run therefore begins with the pinned application configuration, initialized fixtures, and a clean task state. Section~\ref{sec:shared-execution} details the shared mechanisms used for subsequent human and agent execution.

In addition, to reduce false positives from oracle assertions that might otherwise accept a generic or accidentally reached state, \FrameworkName uses task fingerprints as an additional safeguard. We examine every task and apply fingerprints wherever its inputs and expected outcome can naturally incorporate a task-specific marker, resulting in 30 of the 84 tasks. For each such task, the fingerprint generator derives short tokens from a SHA-256 hash of the project and task identifiers. As required by each task, \FrameworkName substitutes these tokens into the initialization script, task fixtures, oracle assertions, and task description. The expected token thus becomes part of both the task instance and its completion criteria, allowing the oracle to verify the intended task-specific result.

\subsubsection{Human--Agent Task Execution}
\label{sec:human-agent-execution}
Fig.~\ref{fig:framework}(b) shows the human user and keyboard agent in corresponding task-execution roles. After provisioning, the TUI is already running inside the Docker sandbox and rendered through the host Tmux session. Both receive the same task description and interact with this running TUI through the three shared stages of \emph{observation}, \emph{interpretation}, and \emph{action}, which correspond to $(o_i,r_i,a_i)$ in Eq.~\ref{eq:interaction-trajectory}.

For the human user, the running TUI first renders the screen. The user views this screen with human eyes, and the human brain interprets it. The user then presses keys as physical key events, which reach the TUI as terminal input. For the keyboard agent, capture pane obtains a TUI screenshot of the same rendered screen. The LLM core updates context from this screenshot and generates an action. Tmux send-keys then issues those keys as virtual key events, which likewise reach the TUI as terminal input. The TUI renders the updated screen for the next observation, and either role continues this loop until submit~/~exit. \FrameworkName records the resulting screen--key trajectory for either role. During this process, \FrameworkName records the timestamp of every physical or virtual key event for later analysis.

\subsubsection{Independent Task Verification}
\label{sec:task-verification}
As shown in Fig.~\ref{fig:framework}(c), once either the human user or keyboard agent submits the task or exits the TUI, \FrameworkName activates the verification engine without sending a further key to the TUI. Following the software-testing notion of a test oracle~\cite{DBLP:journals/tse/BarrHMSY15}, we predefine assertions for each task's expected final state. The engine evaluates three classes of assertions. Shell-state and filesystem assertions query the resulting Docker state through \texttt{docker exec}, whereas screen assertions inspect the final host Tmux pane. The aggregate result is PASS only when every assertion passes and FAIL otherwise.

Fig.~\ref{fig:oracle-primitives} details the six assertion primitives used to compose each task-specific oracle.
\textbf{(1) Shell-state assertions:} \texttt{check\_shell} executes a command in the container and compares its output with the expected output.
\textbf{(2) Filesystem assertions:} \texttt{check\_filesystem} searches configured roots for a file containing the specified path and content tokens. \texttt{check\_filesystem\_content} searches for content without constraining the path.
\textbf{(3) Screen assertions:} \texttt{check\_screen} verifies that specified text appears on the final screen. \texttt{check\_screen\_absent} verifies that specified text does not appear. \texttt{check\_screen\_highlight} verifies that specified text is visually highlighted. Detailed implementations of these checks are available in our repository.

\subsection{Shared Human--Agent Execution Mechanism}
\label{sec:shared-execution}

\begin{wrapfigure}[12]{r}{0.70\textwidth}
    \centering
    \includegraphics[width=\linewidth]{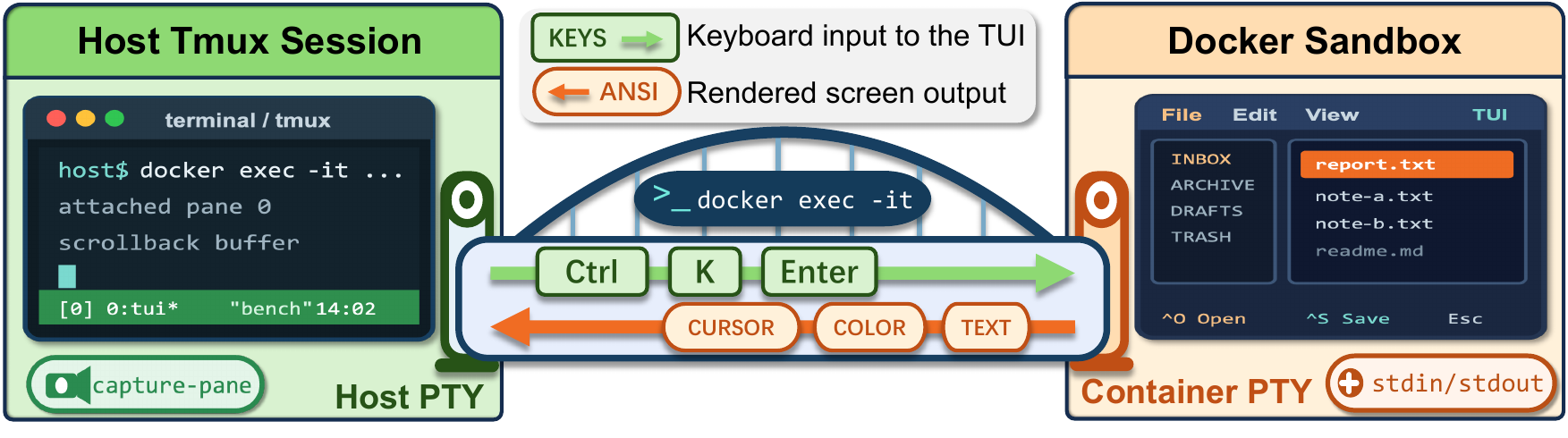}
    \caption{Shared terminal transport between the host Tmux session and Docker sandbox. Keys cross the PTY bridge into the TUI, and ANSI output returns on the same channel for host-side capture.}
    \Description{A suspension-style PTY bridge connects a host Tmux terminal on the left to a Docker sandbox on the right. Orange keycaps travel toward the TUI, while teal ANSI packets return toward Tmux. Host and container PTYs form the bridge endpoints. Capture-pane reads the retained Tmux screen, and stdin/stdout marks the TUI I/O boundary.}
    \label{fig:shared-execution}
\end{wrapfigure}

Human and agent trajectories are comparable only when both actors observe the same rendered screen and act on the same application input boundary.
To enforce this alignment at runtime, \FrameworkName uses the shared terminal transport shown in Fig.~\ref{fig:shared-execution}.

In the \textbf{host Tmux session}, Tmux creates an interactive pane and its pseudo-terminal (PTY). A PTY provides a bidirectional interface that lets a process communicate as if it were attached to a conventional terminal~\cite{linuxmanpages2026pty}. A human enters physical keystrokes into this pane, whereas the agent injects virtual keys with Tmux \texttt{send-keys}.
To connect this pane to the \textbf{Docker sandbox}, \FrameworkName invokes \texttt{docker exec -it}. Docker creates a container-side PTY and attaches it to the TUI's standard input and output. The connected PTYs form the bidirectional bridge shown in Fig.~\ref{fig:shared-execution}. Input from either actor reaches the same TUI stdin. The TUI sends rendered output in the opposite direction as ANSI state containing text, color, and cursor information. Tmux retains the resulting screen in the host pane, where \FrameworkName reads it with \texttt{capture-pane}. Observation and action therefore use the same terminal transport rather than separate paths into the container. The Docker sandbox remains alive after the TUI exits, preserving the final filesystem and shell state for the oracle in Fig.~\ref{fig:framework}(c).

\subsection{Keyboard Agent Design}
\label{sec:keyboard-agent-design}
Section~\ref{sec:shared-execution} places both actors on the same rendered TUI and input boundary. 
We next describe how the keyboard agent implements Eq.~\ref{eq:interaction-trajectory} as a controlled turn-based loop.

\subsubsection{Interaction Loop}
\label{sec:keyboard-agent-loop}
Each turn instantiates one tuple $(o_i,r_i,a_i)$. \textbf{Observation.} The host-side \InteractionController captures the current Tmux screen and encodes it using one of three observation representations. The plain representation preserves the visible terminal characters and layout as unstyled text. The semantic representation augments the same text with XML-like \texttt{<fg:color>} and \texttt{<bg:color>} tags that encode ANSI foreground and background colors. The image representation renders the terminal screen as a PNG image. The controller then adds $o_i$ to the LLM context. \textbf{Interpretation.} The LLM interprets the current observation in light of the interaction so far, and we record the reasoning tokens it produces as $r_i$. \textbf{Action.} The agent must emit exactly one $a_i$ from the four permitted actions: \texttt{type}, \texttt{press}, \texttt{submit}, or \texttt{wait}. The controller rejects any other output as invalid. It executes a valid action and, when the interaction continues, captures the resulting screen to begin the next turn.

Three of these actions align with human behavior at the shared execution boundary. The agent uses \texttt{type} to enter a literal text string and \texttt{press} to issue a functional key or key combination, such as \keycap{Enter} or \keycap{Ctrl}+\keycap{A}. This separation distinguishes textual input from keyboard commands. The agent uses \texttt{submit} to express the same completion decision as a human submission, and the controller intercepts it rather than delivering it to the TUI. When processing may still be underway, \texttt{wait} pauses and captures the screen again without sending another key.

\subsubsection{Execution Scope and Recording}
\label{sec:keyboard-agent-scope}
To keep this loop within the task-execution role, each task uses a fresh LLM context containing only the task description and action protocol. Throughout a run, the agent has no access to source code, the filesystem, process state, or oracle output. The controller dispatches application-bound keyboard input through Tmux and normalizes it into \AtomicKeyEvents using the counting rule in Section~\ref{sec:agent-klm-cost}. The loop ends on submission, TUI or session exit, or a predefined execution limit, after which the actor-independent oracle evaluates the final state. We record the screen observations and actions together with the reasoning-token and atomic key event counts.



\section{Empirical Study Design}
\label{sec:experiment}

\subsection{Human Experiment Setup}
\label{sec:human-study}

We recruit six participants to independently complete all 84 tasks. Each participant has more than five years of programming experience and is proficient in Linux and terminal use. None participated in the repository selection or task construction described in Section~\ref{sec:benchmark-construction}. The six participants therefore approach the benchmark as experienced terminal users without prior knowledge of its task designs or intended interaction paths. Participation was voluntary, and all participants provided informed consent before the study.

We implement a human experiment harness for macOS and Linux that mirrors the \KeyboardAgent setup in task initialization, sandboxed execution, oracle verification, and trajectory logging. The principal difference is observation delivery: participants view the Tmux pane directly in a maximized desktop terminal rather than receiving an encoded screen representation. Mouse input is disabled to preserve the shared keyboard boundary. The harness maps three consecutive \keycap{Ctrl}+\keycap{G} inputs to the \texttt{submit} action defined in Section~\ref{sec:keyboard-agent-loop} and excludes these control inputs from subsequent interaction statistics. All other keystrokes are delivered to the TUI through the physical keyboard, and the harness records a timestamp for every event.

Each task uses an independent initial state. Participants complete each attempt without interruption and may take breaks between tasks. This preserves the continuous trajectory needed to measure human interaction time from the initial screen to submission. Because the \KeyboardAgent begins every task with a fresh LLM context as described in Section~\ref{sec:keyboard-agent-scope}, we shuffle the human task order and avoid grouping tasks from the same repository. This arrangement reduces carryover from recently learned application conventions. Participants first attempt every task once, then revisit unsuccessful tasks after the full first round until all tasks pass. Our subsequent analyses are based primarily on each participant's first attempt completed through the designated \keycap{Ctrl}+\keycap{G} submission. Finally, completing the study takes approximately one to four hours per participant, reflecting variation in familiarity and pace even among experienced terminal users.

\subsection{Agent Experiment Setup: Selected LLMs and Agents}
\label{sec:agent-study}

\subsubsection{Keyboard Agent Settings}
\label{sec:keyboard-agent-settings}
All agent experiments follow the \KeyboardAgent design in Section~\ref{sec:keyboard-agent-design}, with a maximum of 100 interaction turns per run. We use \path{deepseek-v4-pro} and \path{deepseek-v4-flash} for text-based observations, and the image-capable \path{deepseek-v4-flash-vision-exp} for image observations. We abbreviate these three evaluators as Pro, Flash, and Vision, respectively. All \KeyboardAgent runs are conducted in August 2026 through DeepSeek's official API with thinking mode enabled, allowing us to record the returned \ReasoningTokens. We repeat each task three times per setting to improve the statistical reliability of our conclusions.

\subsubsection{TUI Generation Settings}
\label{sec:tui-generation-settings}
To examine the usability of software produced by contemporary coding agents, we select three flagship agent--model configurations: Claude Code using Claude-Opus-5, Codex using GPT-5.6-Sol, and DeepSeek Harness using DeepSeek-V4-Pro. These configurations represent leading coding-agent capabilities available at the time of our study. Each independently generates one counterpart for every reference TUI using the corresponding repository-level prompt from Section~\ref{sec:generation-prompt-construction}, yielding 45 generated applications. We build a Docker image for each application and verify that it can be installed and launched successfully.

\subsection{Metrics}
We characterize TUI usability through task completion and the interaction trajectories required to achieve it. Task oracles verify successful completion, while recorded trajectories quantify \AgentKLM's operation and interpretation costs and reveal how users and agents interact with the TUI.

\subsubsection{Task Completion}
\label{sec:metric-effectiveness}
\textbf{Success rate} ($SR$) measures task completion. An attempt succeeds only if it passes every assertion in the task-specific oracle. For each task, human $SR$ is the mean of the six participants' binary first-attempt outcomes, while agent $SR$ is the mean of three fresh-context run outcomes. Profile-level $SR$ is the mean of these task-level rates across all 84 tasks.

\subsubsection{Interaction Costs}
\label{sec:metric-interaction-costs}
Following Section~\ref{sec:agent-klm-cost}, we operationalize the two interaction-cost dimensions with role-specific measures. For human trajectories, interaction time ($T_H$) serves as a proxy for interpretation demand, and physical atomic key event count ($N_H$) measures operation. For agent trajectories, reasoning token length ($L_A$) and virtual atomic key event count ($N_A$) serve the corresponding roles. $T_H$ spans the initial screen to the first submission, whereas $L_A$ totals reasoning tokens across turns. $N_H$ and $N_A$ follow the shared counting rule in Section~\ref{sec:agent-klm-cost}. In particular, we do not compare the raw values of $T_H$ and $L_A$, but compare their task rankings instead because they use different units.

\subsubsection{Keyboard Trajectory Characteristics}
\label{sec:metric-trajectory-characteristics}
To characterize interaction composition, we group atomic key events by their observable forms into four classes: \textbf{printable} characters and symbols, \textbf{navigation} keys for cursor or focus movement, \textbf{deletion} keys, and \textbf{control} keys for commands and modifier combinations~\cite{leijten2013keystroke,DBLP:conf/oopsla/YoonM11}. We use observable forms because the same printable key may enter text or invoke a command depending on the TUI state. To compare trajectory order, we construct \textbf{raw-key sequences} and \textbf{class sequences}, with the latter mapping events to the four classes and collapsing consecutive repetitions. For each representation, we compute normalized \textbf{Levenshtein} similarity as $1-d_{\mathrm{Lev}}(s_1,s_2)/\max(|s_1|,|s_2|)$ and normalized longest-common-subsequence (\textbf{LCS}) similarity as $2|\mathrm{LCS}(s_1,s_2)|/(|s_1|+|s_2|)$. Together, the two representations and two similarity measures yield four trajectory-similarity indicators.

\subsubsection{Statistical Analysis}
\label{sec:statistical-analysis}
We aggregate observations by task, using success proportions for completion and medians for interaction costs and pairwise trajectory similarities. Correlation analyses use Spearman's $\rho$ for pairwise task rankings and Kendall's $W$ for agreement across configurations. We compute 95\% percentile confidence intervals from 10,000 hierarchical bootstrap resamples at the applicable repository, task, participant, and agent-run levels~\cite{davison1997bootstrap,DBLP:journals/tosem/JapkeGLB25}. Each interval is reported as $[L, U]$, where $L$ and $U$ are its lower and upper bounds.

\begin{table}[t]
\centering
\caption{Human--agent profiles across 84 tasks. P1--P6 are participant-level means. Human mean and agent mean average the 84 task-level completion rates ($SR$) or costs, after within-task aggregation (mean for $SR$ and median for others). Units are percent for $SR$, seconds for $T_H$, thousands of reasoning tokens for $L_A$, and AKE for $N_H$ and $N_A$. The same unit convention applies to Tables~\ref{tab:rq2-main} and~\ref{tab:rq3-master}. Bold rank coefficients denote 95\% bootstrap CIs excluding zero. The same bolding convention applies to Tables~\ref{tab:rq1-trajectories} and~\ref{tab:rq2-main}.}
\label{tab:rq1-alignment}
\footnotesize
\setlength{\tabcolsep}{2.0pt}
\newcommand{\ghead}[2]{%
  \bgroup
  \setbox0=\hbox{\textbf{#1}}%
  \copy0
  \llap{\raisebox{-0.62ex}[0pt][0pt]{\hbox to \wd0{\hss\rule{#2}{\cmidrulewidth}\hss}}}%
  \egroup}
\newcolumntype{E}{wc{2.2em}}
\begin{tabular}{l!{\hspace{3pt}}*{6}{E!{\hspace{3pt}}}c!{\hspace{3pt}}c!{\hspace{3pt}}c!{\hspace{3pt}}c}
\toprule
\rowcolor{gray!15}\multicolumn{1}{l!{\hspace{3pt}}}{} & \multicolumn{7}{c!{\hspace{3pt}}}{\ghead{Human}{20em}} & \multicolumn{1}{c!{\hspace{3pt}}}{\ghead{Agent}{2.6em}} & \multicolumn{2}{c}{\ghead{Rank correlation}{9.8em}} \\
\rowcolor{gray!15}\multicolumn{1}{c!{\hspace{3pt}}}{\multirow{-2}{*}{\textbf{Measure}}} & \textbf{P1} & \textbf{P2} & \textbf{P3} & \textbf{P4} & \textbf{P5} & \textbf{P6} & \textbf{Mean} & \textbf{Mean} & $\boldsymbol{\rho}$ & \textbf{95\% CI} \\
\midrule
\rowcolor{blue!10}\textbf{\makebox[8.3em][l]{Completion} ($SR$)} & 78.57 & 61.90 & 83.33 & 85.71 & 86.90 & 61.90 & 76.39 & 82.94 & 0.3476 & [-0.0283, 0.5675] \\
\addlinespace[1pt]
\rowcolor{green!10}\textbf{\makebox[6.8em][l]{Interpretation} ($T_H$/$L_A$)} & 206.1 & 61.39 & 32.07 & 72.14 & 103.7 & 45.04 & 57.32 & 18.29 & \textbf{0.6219} & [0.2036, 0.7477] \\
\addlinespace[1pt]
\rowcolor{orange!14}\textbf{\makebox[6.3em][l]{Operation} ($N_H$/$N_A$)} & 49.50 & 46.33 & 34.83 & 59.98 & 58.71 & 71.79 & 43.45 & 34.05 & \textbf{0.7687} & [0.4603, 0.8847] \\
\rowcolor{orange!5}\quad$\hookrightarrow$ Printable (\emph{e.g.}, \keycap{A}) & 19.54 & 18.07 & 18.98 & 25.48 & 24.54 & 30.13 & 19.71 & 25.48 & \textbf{0.8939} & [0.6498, 0.9372] \\
\rowcolor{orange!5}\quad$\hookrightarrow$ Navigation (\emph{e.g.}, \keycap{$\uparrow$}) & 22.82 & 21.69 & 10.05 & 26.54 & 26.90 & 28.19 & 15.99 & 4.524 & \textbf{0.4760} & [0.0359, 0.7129] \\
\rowcolor{orange!5}\quad$\hookrightarrow$ Deletion (\emph{e.g.}, \keycap{Delete}) & 3.857 & 3.214 & 3.000 & 4.524 & 3.214 & 8.381 & 2.940 & 0.250 & \textbf{0.3231} & [0.0183, 0.5051] \\
\rowcolor{orange!5}\quad$\hookrightarrow$ Control (\emph{e.g.}, \keycap{Enter}) & 3.286 & 3.357 & 2.810 & 3.440 & 4.060 & 5.083 & 2.911 & 3.524 & \textbf{0.6452} & [0.2386, 0.8168] \\
\rowcolor{orange!5}\quad$\hookrightarrow$ Non-text aggregate & 29.96 & 28.26 & 15.86 & 34.50 & 34.18 & 41.65 & 23.74 & 8.571 & \textbf{0.4813} & [0.0629, 0.7077] \\
\bottomrule
\end{tabular}
\end{table}

\subsection{Research Questions}
To systematically evaluate \AgentAsUser, we formulate the following three research questions (RQs). They progress from human--agent correspondence, through the effect of observation on that correspondence, to the task-based usability of generated TUIs:
\begin{description}
    \item[\textbf{RQ1.}] How do agent completion, the two interaction costs, and the resulting paths correspond to human interaction on the same tasks?
    \item[\textbf{RQ2.}] How do plain, semantic, and image observations affect that correspondence, given that humans see the rendered screen and the agent receives a discrete representation?
    \item[\textbf{RQ3.}] What is the task-based usability of TUIs generated by three leading coding agents, relative to the reference TUIs?
\end{description}

\section{Results}
\label{sec:results}

\definecolor{findingRQOne}{HTML}{2A7F7A}
\definecolor{findingRQTwo}{HTML}{426DA9}
\definecolor{findingRQThree}{HTML}{B66A2C}

\newcommand{\rqconclusion}[3]{%
    \begin{center}
    \begingroup
    \setlength{\fboxsep}{6pt}%
    \begin{tikzpicture}
        \node[
            draw=#3!70!black,
            fill=#3!7,
            rounded corners=3pt,
            line width=\fboxrule,
            inner sep=\fboxsep,
            text width=\dimexpr\linewidth-2\fboxsep-2\fboxrule\relax,
            align=left
        ] {\small\textcolor{#3!80!black}{\textbf{#1}} \textit{#2}};
    \end{tikzpicture}%
    \endgroup
    \end{center}
}

Our analysis covers 3,528 agent attempts and \RQOneHumanTaskRuns{} human first attempts from \RQOneHumanN{} participants across 84 tasks. To answer RQ1, we compare human and agent evaluations of the reference TUIs to identify similarities and differences in completion, the two \AgentKLM costs, and interaction paths. To answer RQ2, we examine how different visual observation encodings affect those measurements across five observation--evaluator configurations. To answer RQ3, we compare each reference TUI with its three independently generated counterparts using the same completion, interpretation, and operation measures.

\subsection{RQ1: Human--Agent Behavioral Relationship}
\label{sec:rq1-results}

\noindent\textbf{Success rate alone cannot establish human--agent usability correspondence.}
Table~\ref{tab:rq1-alignment} summarizes the RQ1 results using \texttt{deepseek-v4-pro} with plain observations as a representative text-based agent configuration. Participants complete \RQOneHumanFirstPasses{} of \RQOneHumanTaskRuns{} first attempts (\RQOneHumanFirstPass{}), while the agent completes \RQOneAgentSuccesses{} of \RQOneAgentAttempts{} independent fresh-context runs (\RQOneAgentPass{}). Viewed only in aggregate, these rates appear similar. However, their task-level correlation is limited and uncertain ($\rho=\RQOneSuccessRho$, 95\% CI [\RQOneSuccessRhoLow, \RQOneSuccessRhoHigh]). Five tasks have high agent but low human completion rates, while six show the reverse. Overestimation on some tasks and underestimation on others can therefore cancel in the aggregate, making two evaluations appear more closely aligned than they are task by task.

This limitation is also inherent to the metric. Success rate collapses an entire trajectory into a final pass-or-fail outcome, treating direct completion and completion after extensive exploration alike. It remains informative about whether tasks are completed, but cannot by itself characterize the usability demands encountered along the way. We therefore examine the interaction cost introduced in Section~\ref{sec:agent-klm-cost} through its \InterpretationCost and \OperationCost dimensions.

\begin{table}[t]
\centering
\caption{Trajectory similarity across 84 tasks. H--H, H--A, and A--A denote human--human, human--agent, and agent--agent pairings. Pairwise values are task-level median similarities multiplied by 100. Each subscripted $\rho_{X,Y}$ is Spearman's rank correlation between the task-level similarities for pairing types $X$ and $Y$.}
\label{tab:rq1-trajectories}
\scriptsize
\setlength{\tabcolsep}{1pt}
\newcommand{\ghead}[2]{%
  \bgroup
  \setbox0=\hbox{\textbf{#1}}%
  \copy0
  \llap{\raisebox{-0.62ex}[0pt][0pt]{\hbox to \wd0{\hss\rule{#2}{\cmidrulewidth}\hss}}}%
  \egroup}
\newcommand{\simcell}[1]{\makebox[2.4em][c]{#1}}
\begin{tabular}{c!{\hspace{2.8pt}}c!{\hspace{2.8pt}}c!{\hspace{2.8pt}}c!{\hspace{2.8pt}}c!{\hspace{2.8pt}}c!{\hspace{2.8pt}}c!{\hspace{2.8pt}}c!{\hspace{2.8pt}}c!{\hspace{2.8pt}}c!{\hspace{2.8pt}}c}
\toprule
\rowcolor{gray!15} & &
\multicolumn{3}{c!{\hspace{2.8pt}}}{\ghead{Pairwise similarity}{8em}} &
\multicolumn{6}{c}{\ghead{Rank correlation}{34em}} \\
\rowcolor{gray!15}\multirow{-2}{*}{\textbf{Measure}} &
\multirow{-2}{*}{\textbf{Sequence}} &
\simcell{\textbf{H--H}} & \simcell{\textbf{H--A}} & \simcell{\textbf{A--A}} &
$\boldsymbol{\rho}_{\mathrm{H\text{--}H,H\text{--}A}}$ & \textbf{95\% CI} &
$\boldsymbol{\rho}_{\mathrm{H\text{--}H,A\text{--}A}}$ & \textbf{95\% CI} &
$\boldsymbol{\rho}_{\mathrm{H\text{--}A,A\text{--}A}}$ & \textbf{95\% CI} \\
\midrule
\rowcolor{blue!10} & Raw-key & \simcell{28.57} & \simcell{34.07} & \simcell{73.86} &
\textbf{0.5224} & [0.2096, 0.7292] &
$-$0.1267 & [$-$0.4147, 0.1703] &
0.2651 & [$-$0.1133, 0.5905] \\
\rowcolor{blue!10}\multirow{-2}{*}{\textbf{Levenshtein}} & Class & \simcell{49.56} & \simcell{44.95} & \simcell{75.00} &
\textbf{0.5554} & [0.2842, 0.7423] &
0.1179 & [$-$0.2707, 0.4480] &
0.3147 & [$-$0.0150, 0.5718] \\
\addlinespace[1pt]
\rowcolor{green!10} & Raw-key & \simcell{43.56} & \simcell{50.05} & \simcell{82.89} &
\textbf{0.5558} & [0.2325, 0.7678] &
$-$0.1313 & [$-$0.4101, 0.1565] &
0.2466 & [$-$0.1154, 0.5624] \\
\rowcolor{green!10}\multirow{-2}{*}{\textbf{LCS}} & Class & \simcell{62.73} & \simcell{60.08} & \simcell{85.71} &
\textbf{0.6340} & [0.3846, 0.7850] &
0.0992 & [$-$0.2583, 0.4123] &
0.2715 & [$-$0.0851, 0.5458] \\
\bottomrule
\end{tabular}
\end{table}

\noindent
\textbf{Interaction costs reveal shared demand beneath different outcomes.}
For the cost comparisons, each task is represented by the median of the six participants' first attempts and the median of the agent's three fresh runs. This aggregation includes all \RQOneComparisonTasks{} tasks and reduces the influence of occasional extreme trajectories. The task-level human interaction time ($T_H$) and agent reasoning-token count ($L_A$) have a rank correlation of $\rho=\RQOneReasoningRho$ (95\% CI [\RQOneReasoningRhoLow, \RQOneReasoningRhoHigh]). Tasks with longer human interaction times also tend to consume more agent reasoning. Although $T_H$ and $L_A$ use different units and do not measure equal cognitive effort, their shared ordering is consistent with a common task-level interpretation demand.

The correspondence is stronger for \OperationCost. Physical and virtual \AtomicKeyEvents ($N_H$ and $N_A$) have a rank correlation of $\rho=\RQOneCostRho$ (95\% CI [\RQOneCostRhoLow, \RQOneCostRhoHigh]). Both measures observe actions at the same keyboard boundary, and their stronger alignment shows that tasks requiring more human keyboard work generally also require more agent interaction. Consequently, human and agent outcomes may diverge while the amount of interpretation and operation demanded by the tasks remains related. These graded costs capture a human--agent relationship that the binary success outcome alone does not expose.

\rqconclusion{Finding 1.}{Similar success rates can hide differences in which tasks humans and the agent complete. In contrast, human and agent costs follow similar patterns across tasks, especially for operation.}{findingRQOne}

\noindent
\textbf{Aligned operation costs conceal different interaction habits.}
The overall \OperationCost relationship could have a simple mechanical explanation: tasks requiring longer literal input would increase both human and agent key counts. To test whether text entry alone drives the result, we decompose \AtomicKeyEvents into the four analytical key classes defined in Section~\ref{sec:metric-trajectory-characteristics}. Table~\ref{tab:rq1-alignment} shows positive task-level correlations for every class, with all 95\% CIs excluding zero. Printable input has the strongest relationship ($\rho=\RQOnePrintableRho$), but control ($\rho=0.65$), navigation ($\rho=0.48$), and deletion ($\rho=0.32$) remain related as well. Consistently, aggregating all non-printable events retains a correlation of $\rho=\RQOneNontextRho$ (95\% CI [\RQOneNontextRhoLow, \RQOneNontextRhoHigh]). The operation relationship is therefore not specific to one kind of key: even deletion, despite being rare in agent paths, retains a positive association.

\begin{wrapfigure}[10]{r}{0.52\textwidth}
    \centering
    \includegraphics[width=\linewidth]{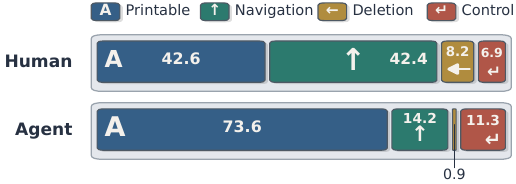}
    \caption{Human and agent key-class composition. Key width is the share of atomic key events (\%).}
    \Description{Two keyboard strips compare printable, navigation, deletion, and control key shares. Human input splits mainly between printable and navigation keys, while agent input is predominantly printable.}
    \label{fig:rq1-key-composition}
\end{wrapfigure}

However, task-level correlation does not imply identical interaction habits. Fig.~\ref{fig:rq1-key-composition} shows that printable keys account for \RQOneAgentPrintableShare{} of agent actions but only \RQOneHumanPrintableShare{} of human actions. In contrast, participants devote \RQOneHumanNavigationShare{} of their input to navigation, compared with \RQOneAgentNavigationShare{} for the agent, and also use more deletion (\RQOneHumanDeletionShare{} versus \RQOneAgentDeletionShare{}). This pattern is consistent with greater agent reliance on literal input and less incremental navigation and correction. Thus, humans and the agent can rank task operation demand similarly while using different key mixtures.

\noindent
\textbf{Sequence-level alignment increases with human path consistency.}
Table~\ref{tab:rq1-trajectories} extends Table~\ref{tab:rq1-alignment} from aggregate costs to pairwise path similarity. For each task and pairing type, we take the median similarity before summarizing all 84 tasks. Across the four measures, the median human--agent minus human--human differences range from $-0.03$ to $+0.05$. These small gaps do not consistently favor either pairing. Their 95\% CIs all include zero, so the data show no consistent human--agent similarity deficit. On the 0--100 scale used in Table~\ref{tab:rq1-trajectories}, human--agent versus human--human similarity is 34.07 versus 28.57 for raw-key Levenshtein and 60.08 versus 62.73 for class-sequence LCS. In comparison, agent--agent similarity ranges from 73.86 to 85.71 and exceeds both other pairings in every row. Repeated agent runs therefore follow a more regular path.

Across the 84 tasks, we further calculate Spearman's rank correlations among the three pairing types, as reported in Table~\ref{tab:rq1-trajectories}. Across all four sequence measures, only human--human and human--agent similarities covary reliably across tasks ($\rho_{\mathrm{H\text{--}H,\,H\text{--}A}}=0.52$--$0.63$), with all four 95\% CIs excluding zero. Neither human--human nor human--agent similarity correlates significantly with agent--agent similarity: every $\rho_{\mathrm{H\text{--}H,\,A\text{--}A}}$ and $\rho_{\mathrm{H\text{--}A,\,A\text{--}A}}$ interval includes zero. This contrast suggests that human--agent path alignment is associated with how consistently humans approach a task. Although the agent's own repeatability is high in absolute terms, it does not explain which tasks elicit more human-like agent paths.

\rqconclusion{Finding 2.}{Human and agent operation costs rise together across key classes despite different key mixtures. Human--agent path similarity is close to human--human similarity and increases when humans follow more consistent routes, while repeated agent runs remain more regular.}{findingRQOne}

Taken together, RQ1 suggests that the agent is better viewed as a measure of task-level interaction demand in \AgentAsUser. It is not a direct replica of human behavior. Completion alone misses this relationship. Cost analysis shows that the same tasks tend to require more operations from both humans and the agent. Trajectory analysis shows that agent paths align more closely with human paths when humans follow more consistent routes. 

\subsection{RQ2: Effects of Observation Representations}
\label{sec:rq2-results}

\begin{figure}[t]
    \centering
    \includegraphics[width=\textwidth]{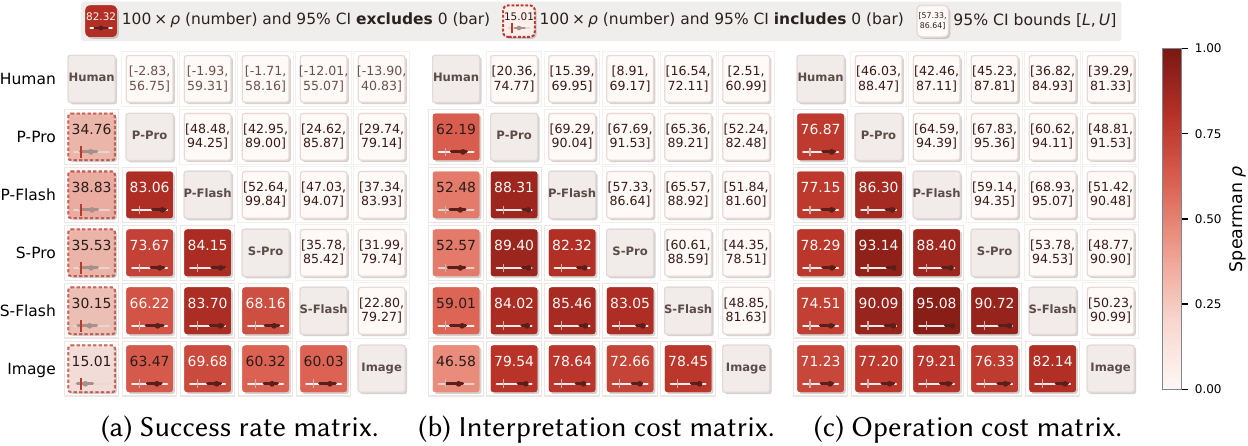}
    \caption{Task-level Spearman rank correlations among humans and five agent configurations across 84 reference tasks. P and S denote plain and semantic observations, and Image denotes the image--Vision configuration. Color encodes $\rho$. The lower triangle shows $100\times\rho$ and 95\% CI bars, and the upper triangle gives $[L,U]$. Dashed borders denote CIs spanning zero, with red ticks marking zero.}
    \Description{Three Spearman correlation matrices for completion ($SR$), interpretation ($T_H$/$L_A$), and operation (AKE), comparing humans with five agent configurations.}
    \label{fig:rq2-sensitivity}
\end{figure}

\noindent\textbf{Richer observation representations provide no consistent evaluation advantage.}
As shown in Fig.~\ref{fig:framework}(b), one major difference between the human participant and the keyboard agent lies in the observation path. The human participant views the rendered pane continuously, whereas the keyboard agent receives a discrete representation. We compare plain text, semantic text with ANSI color cues, and the rendered image, whose information richness follows the intuitive ordering plain \(<\) semantic \(<\) image. Intuitively, richer input should make interface cues easier to identify and reduce the reasoning and keyboard work required for the next action. To examine whether this expectation holds, we keep the tasks, fixtures, action boundary, and oracles fixed.

\begin{table}[t]
\centering
\caption{Human and observation--evaluator profiles across 84 tasks. Cost cells are means of 84 task-level medians. $W$ is Kendall concordance among five agent configurations.}
\label{tab:rq2-main}
\footnotesize
\setlength{\tabcolsep}{2.4pt}
\newcommand{\ghead}[2]{%
  \bgroup
  \setbox0=\hbox{\textbf{#1}}%
  \copy0
  \llap{\raisebox{-0.62ex}[0pt][0pt]{\hbox to \wd0{\hss\rule{#2}{\cmidrulewidth}\hss}}}%
  \egroup}
\begin{tabular}{l!{\hspace{3pt}}c!{\hspace{3pt}}c!{\hspace{3pt}}c!{\hspace{3pt}}c!{\hspace{3pt}}c!{\hspace{3pt}}c!{\hspace{3pt}}c!{\hspace{3pt}}c}
\toprule
\rowcolor{gray!15} & & \multicolumn{2}{c!{\hspace{3pt}}}{\ghead{Plain}{4.8em}} & \multicolumn{2}{c!{\hspace{3pt}}}{\ghead{Semantic}{4.8em}} & \multicolumn{1}{c!{\hspace{3pt}}}{\ghead{Image}{2.6em}} & \multicolumn{2}{c}{\ghead{Concordance}{9em}} \\
\rowcolor{gray!15}\multirow{-2}{11em}{\centering\textbf{Measure}} & \multirow{-2}{*}{\textbf{Human}} & \textbf{Pro} & \textbf{Flash} & \textbf{Pro} & \textbf{Flash} & \textbf{Vision} & $\boldsymbol{W}$ & \textbf{95\% CI} \\
\midrule
\rowcolor{blue!10}\textbf{\makebox[8.5em][l]{Completion} ($SR$)} & 76.39 & 82.94 & 83.33 & 84.13 & 84.13 & 68.25 & \textbf{0.7606} & [0.5366, 0.8566] \\
\addlinespace[1pt]
\rowcolor{green!10}\textbf{\makebox[7em][l]{Interpretation} ($T_H$/$L_A$)} & 57.32 & 18.29 & 23.54 & 15.03 & 16.64 & 22.95 & \textbf{0.8575} & [0.7592, 0.9078] \\
\addlinespace[1pt]
\rowcolor{orange!14}\textbf{\makebox[6.5em][l]{Operation} ($N_H$/$N_A$)} & 43.45 & 34.05 & 42.17 & 32.54 & 31.94 & 34.49 & \textbf{0.8869} & [0.7592, 0.9594] \\
\bottomrule
\end{tabular}
\end{table}

Table~\ref{tab:rq2-main} reports the resulting profiles. The four text configurations have similar success rates. At the profile level, semantic observations yield lower mean costs than plain observations under both evaluators, although all paired 95\% bootstrap intervals include zero. The image--Vision configuration has lower completion than the text configurations, but it changes both the observation modality and the evaluator, so the difference cannot be attributed to image input alone. Across the five configurations, all three measures show high task-ranking concordance (\(W=0.76\)--\(0.89\)), with the strongest agreement for the two cost dimensions. Thus, although the configurations shift absolute values, they yield largely consistent task rankings for each measure used in the usability evaluation.

\noindent
\textbf{Observation representations preserve the core human--agent relationships.}
To investigate more deeply whether the relationships observed in RQ1 change with the observation representation, we calculate correlation matrices for humans and the five agent configurations on completion, interpretation, and operation, as shown in Fig.~\ref{fig:rq2-sensitivity}.
A remarkably consistent human--agent pattern emerges across all five configurations. Completion provides no significant correspondence with humans under any configuration. In contrast, interpretation retains a moderate relationship and operation retains the strongest relationship, with all ten cost intervals excluding zero. This ordering reproduces the central RQ1 pattern without depending on the plain-text configuration used there.

The agent configurations are also positively associated with one another on every measure, with all pairwise intervals excluding zero. Although observation--evaluator choices shift the absolute profiles in Table~\ref{tab:rq2-main}, they preserve the relative pattern of measurements across tasks and the ordering from completion to interpretation to operation. In practice, fixing one configuration therefore provides a stable basis for comparative usability evaluation, provided that absolute values are interpreted within that configuration. This consistency also supports the robustness of the \AgentAsUser framework to the tested observation representations.
\rqconclusion{Finding 3.}{Richer observations provide no consistent advantage. Across all five configurations, interpretation and operation correlate significantly with humans, whereas completion does not. \AgentAsUser remains stable across the tested observation representations.}{findingRQTwo}

\subsection{RQ3: Application to Agent-Generated TUIs}
\label{sec:rq3-results}
\begin{table}[t]
\centering
\caption{Reference and generated-TUI profiles across 84 tasks. Means are taken over task-level summaries, and costs use three-run medians. Changes from the reference are shown in parentheses. Green marks an improvement, red a worse result, and bold the best generated implementation in each configuration.}
\label{tab:rq3-master}
\footnotesize
\setlength{\tabcolsep}{1.8pt}
\definecolor{deltaBetter}{HTML}{0F6B3C}
\definecolor{deltaWorse}{HTML}{C0392B}
\newlength{\rqThreeCol}
\settowidth{\rqThreeCol}{\textbf{DeepSeek Harness}}
\newlength{\rqThreeHead}
\settowidth{\rqThreeHead}{\textbf{+ DeepSeek-V4-Pro}}
\ifdim\rqThreeHead>\rqThreeCol
  \setlength{\rqThreeCol}{\rqThreeHead}
\fi
\newcommand{\twolinehead}[2]{%
  \makebox[\rqThreeCol][c]{\begin{tabular}{@{}c@{}}\textbf{#1}\\\textbf{#2}\end{tabular}}%
}
\newcommand{\vchead}[1]{\raisebox{-0.6\baselineskip}{\textbf{#1}}}
\newlength{\rqThreeVal}
\newlength{\rqThreeDelta}
\settowidth{\rqThreeVal}{\textbf{85.71}}
\settowidth{\rqThreeDelta}{(+15.20)}
\newcommand{\up}[2]{\makebox[\rqThreeVal][l]{#1}\,\makebox[\rqThreeDelta][r]{\textcolor{deltaBetter}{(#2)}}}
\newcommand{\dn}[2]{\makebox[\rqThreeVal][l]{#1}\,\makebox[\rqThreeDelta][r]{\textcolor{deltaWorse}{(#2)}}}
\newcommand{\eq}[1]{\makebox[\rqThreeVal][l]{#1}\,\makebox[\rqThreeDelta][r]{(0)}}
\begin{tabular}{l!{\hspace{1.3pt}}c!{\hspace{1.3pt}}c!{\hspace{1.3pt}}c!{\hspace{1.3pt}}c!{\hspace{1.3pt}}c}
\toprule
\rowcolor{gray!15}
\vchead{\makebox[9em][c]{Measure}} &
\vchead{Configuration} &
\vchead{Reference} &
\twolinehead{Claude Code}{+ Claude-Opus-5} &
\twolinehead{Codex}{+ GPT-5.6-Sol} &
\twolinehead{DeepSeek Harness}{+ DeepSeek-V4-Pro} \\
\midrule
\rowcolor{blue!10} & Plain--Pro & 82.94 & \eq{\textbf{82.94}} & \dn{71.03}{-11.91} & \dn{75.00}{-7.94} \\
\rowcolor{blue!10} & Semantic--Pro & 84.13 & \up{\textbf{85.71}}{+1.58} & \dn{72.22}{-11.91} & \dn{75.40}{-8.73} \\
\rowcolor{blue!10}\multirow{-3}{*}{\textbf{\makebox[6.7em][l]{Completion} ($SR$)}} & Image--Vision & 68.25 & \up{\textbf{74.60}}{+6.35} & \up{69.44}{+1.19} & \up{69.44}{+1.19} \\
\addlinespace[1pt]
\rowcolor{green!10} & Plain--Pro & 18.29 & \dn{\textbf{18.50}}{+0.21} & \dn{26.16}{+7.87} & \dn{21.09}{+2.80} \\
\rowcolor{green!10} & Semantic--Pro & 15.03 & \dn{\textbf{19.15}}{+4.12} & \dn{21.50}{+6.47} & \dn{21.64}{+6.61} \\
\rowcolor{green!10}\multirow{-3}{*}{\textbf{\makebox[6.7em][l]{Interpretation} ($L_A$)}} & Image--Vision & 22.95 & \dn{\textbf{28.79}}{+5.84} & \dn{38.15}{+15.20} & \dn{35.52}{+12.57} \\
\addlinespace[1pt]
\rowcolor{orange!14} & Plain--Pro & 34.05 & \dn{34.21}{+0.16} & \up{\textbf{31.19}}{-2.86} & \up{31.64}{-2.41} \\
\rowcolor{orange!14} & Semantic--Pro & 32.54 & \dn{36.32}{+3.78} & \up{\textbf{27.58}}{-4.96} & \dn{38.24}{+5.70} \\
\rowcolor{orange!14}\multirow{-3}{*}{\textbf{\makebox[6.5em][l]{Operation} ($N_A$)}} & Image--Vision & 34.49 & \dn{\textbf{36.85}}{+2.36} & \dn{42.23}{+7.74} & \dn{40.14}{+5.65} \\
\bottomrule
\end{tabular}
\end{table}

\noindent\textbf{Functional correctness does not ensure usable generated TUIs.}
To apply the relationships established in RQ1 and RQ2, we compare each reference TUI with counterparts produced by three complete generation stacks: DeepSeek Harness using DeepSeek-V4-Pro, Claude Code using Claude-Opus-5, and Codex using GPT-5.6-Sol. All four implementations receive the same 84 tasks, fixtures, agent settings, and oracles. Table~\ref{tab:rq3-master} reports Plain--Pro, Semantic--Pro, and Image--Vision separately. Within each configuration and implementation, completion is the mean of three fresh runs. Interpretation and operation are the means of the 84 task-level three-run medians, using all runs independently of completion.

Table~\ref{tab:rq3-master} shows that the generated implementations achieve completion rates of 69.44\% to 85.71\% across the three evaluation configurations. Here, $SR$ serves as an operational indicator of functional correctness by recording whether a generated TUI can be used to reach the required task state. Interpretation cost and operation cost further quantify the interaction effort accumulated during task execution. All nine generation--configuration profiles have a higher mean \InterpretationCost than the corresponding reference. The \OperationCost comparison is mixed: three generated profiles are lower than the reference, whereas six are higher. The lower values occur for DeepSeek Harness with Plain--Pro and for Codex with Plain--Pro and Semantic--Pro. No generated implementation outperforms the reference across all three measures in any evaluation configuration. Thus, reaching correct task outcomes does not ensure usability: the generated implementations consistently require more interpretation and show no consistent operation advantage.

\noindent
\textbf{Claude Code matches the reference in completion but not in usability.}
Claude Code using Claude-Opus-5 is the only generation stack that matches or exceeds reference completion in all three configurations. It ties the reference under Plain--Pro and exceeds it by 1.58\% and 6.35\% under Semantic--Pro and Image--Vision, respectively. It also has the lowest \InterpretationCost among the generated TUIs in every configuration, making it the strongest generated profile on outcome and interpretation.
Even this profile remains less interaction-efficient than the reference. Its \InterpretationCost exceeds the reference by 0.21k to 5.84k reasoning tokens, and its \OperationCost exceeds the reference by 0.16 to 3.78 AKE across the three configurations. Claude Code therefore matches the reference in completion but not in usability.
Codex further demonstrates that fewer actions do not necessarily indicate better usability. Under Semantic--Pro, it requires fewer actions than Claude Code but achieves lower completion and incurs greater \InterpretationCost.

\rqconclusion{Finding 4.}{Under \AgentAsUser evaluation, TUIs generated by current leading agent frameworks still need better task-based usability. Claude Code using Claude-Opus-5 shows the strongest overall profile, although no generation stack consistently minimizes operation cost.}{findingRQThree}
\begin{figure*}[t]
    \centering
    \begin{subfigure}[t]{0.326\textwidth}
        \centering
        \includegraphics[width=\linewidth]{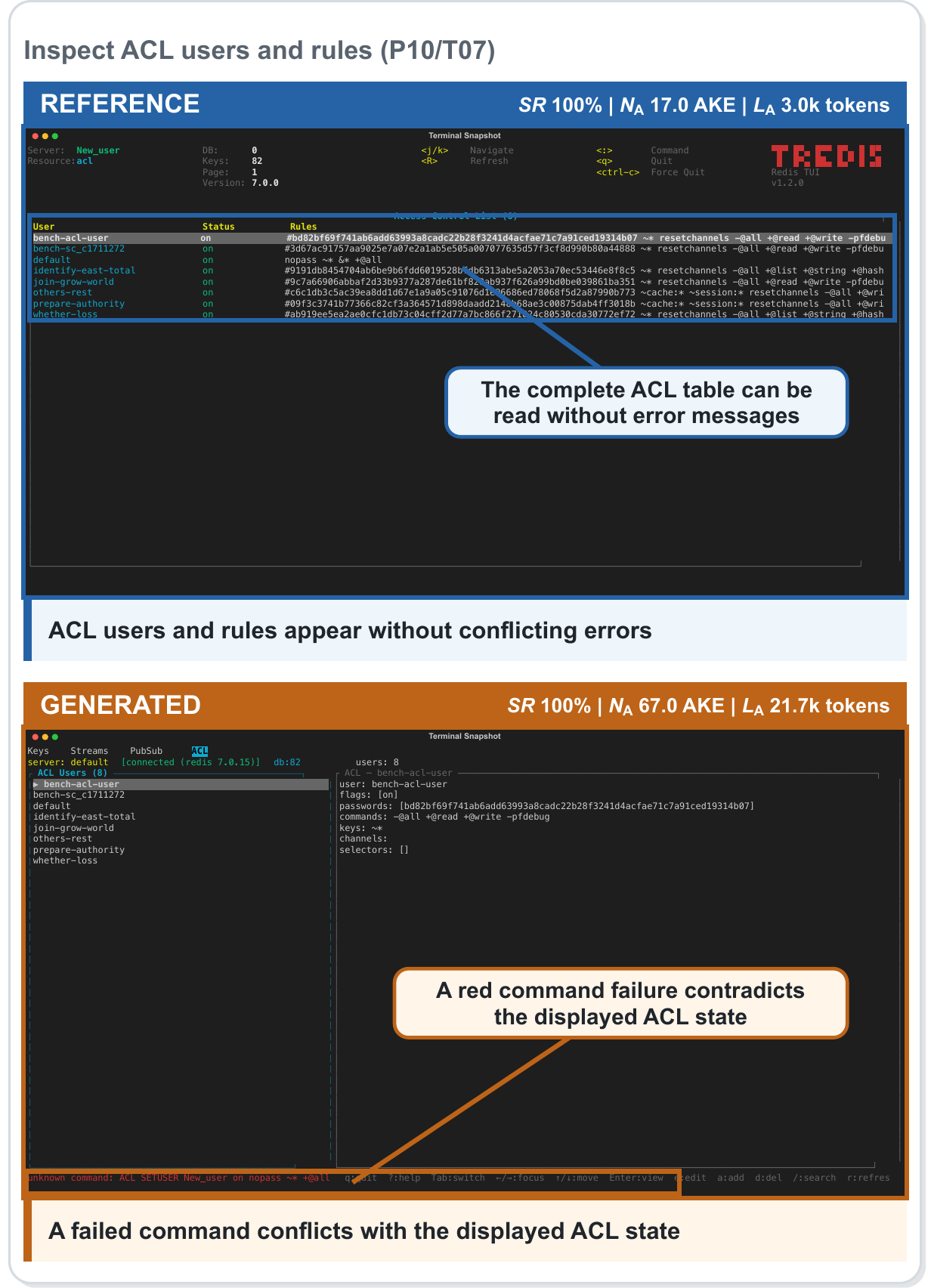}
        \caption{Information Visibility.}
    \end{subfigure}\hfill
    \begin{subfigure}[t]{0.326\textwidth}
        \centering
        \includegraphics[width=\linewidth]{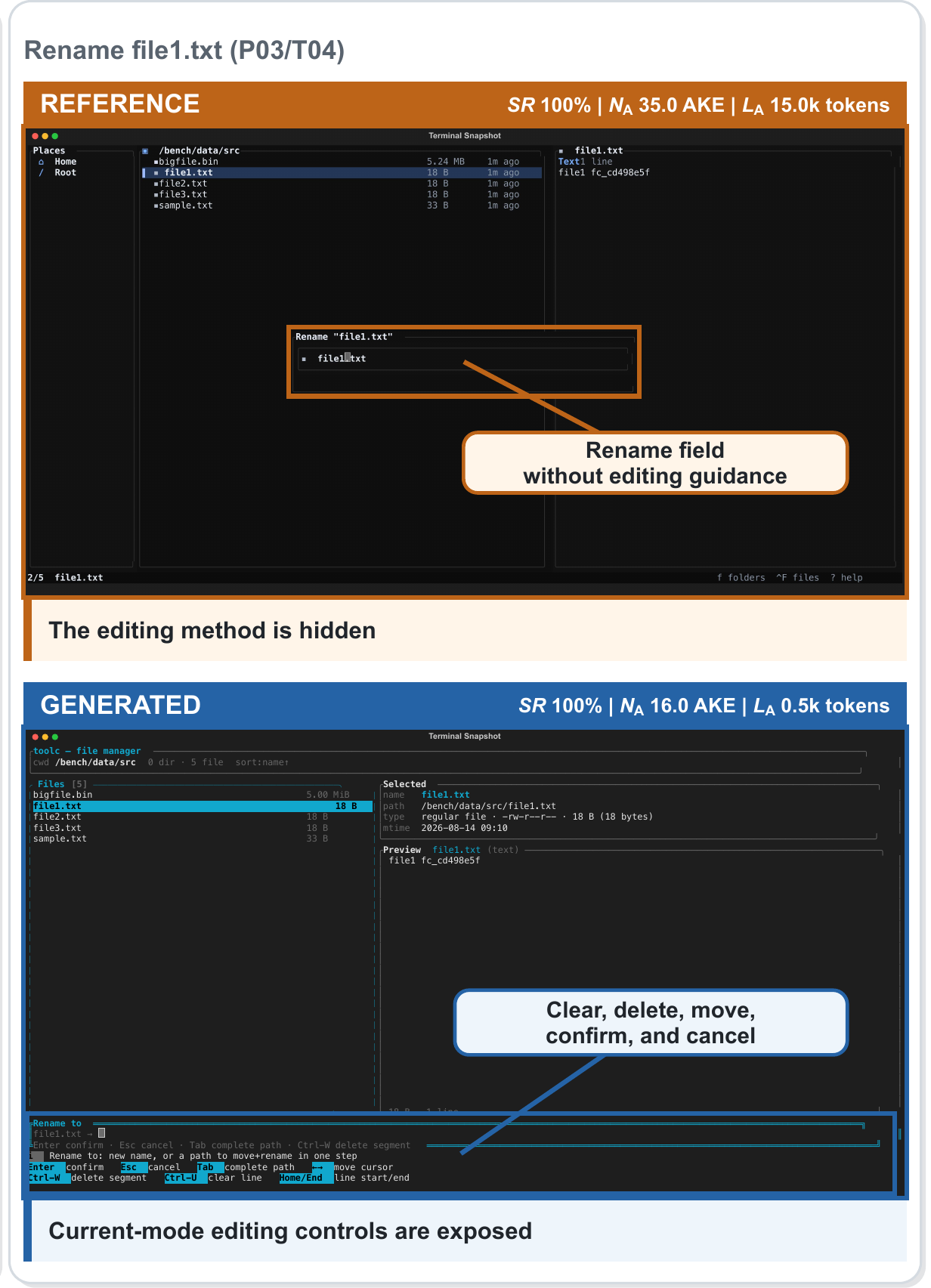}
        \caption{Action Discoverability.}
    \end{subfigure}\hfill
    \begin{subfigure}[t]{0.326\textwidth}
        \centering
        \includegraphics[width=\linewidth]{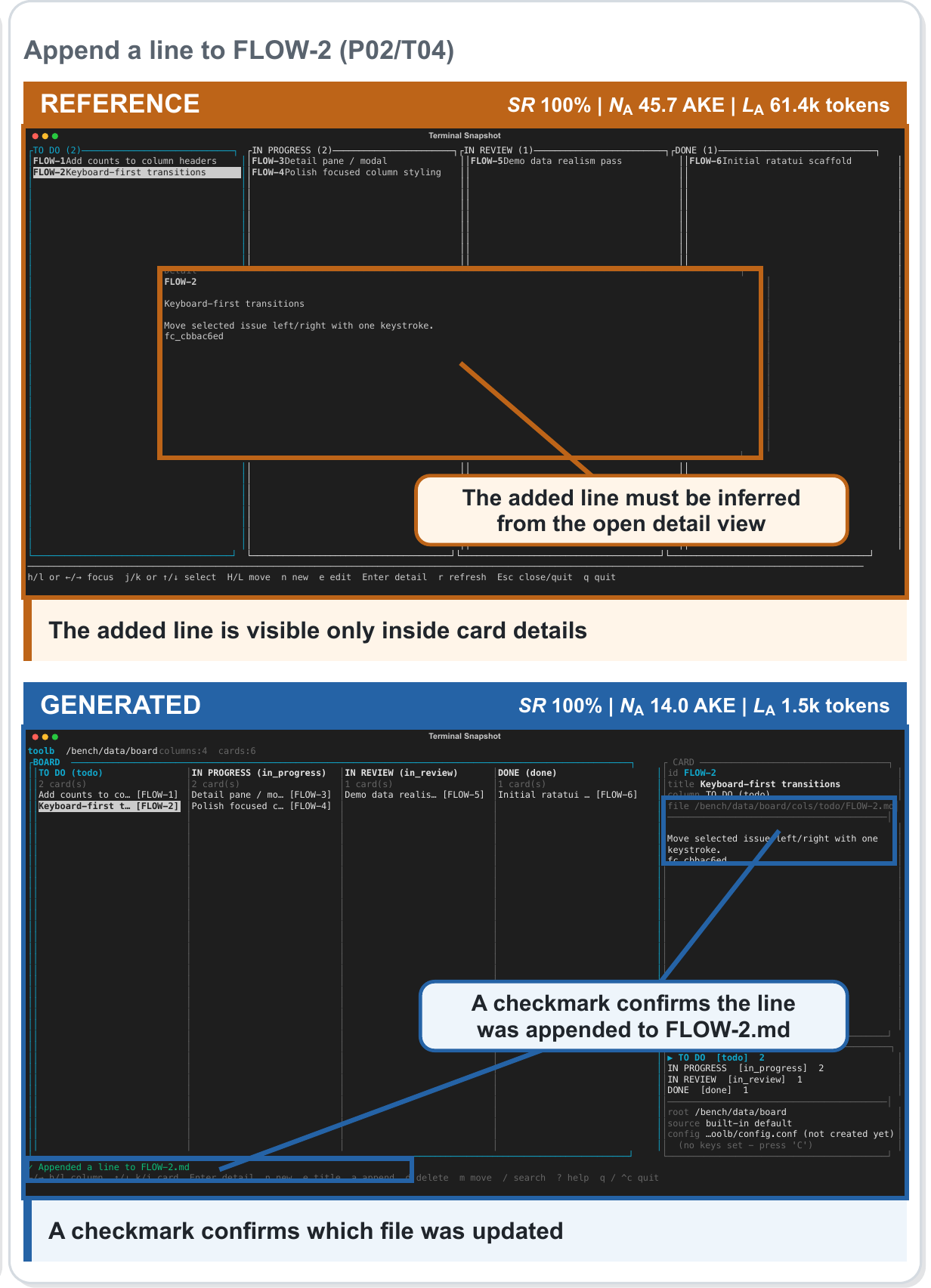}
        \caption{Outcome Feedback.}
    \end{subfigure}
    \caption{Paired Semantic--Pro cases for three TUI design dimensions. Blue marks more usable designs and orange less usable ones. Headers report $SR$, $N_A$ in AKE, and $L_A$ in thousands of reasoning tokens.}
    \Description{Three paired comparisons show complete terminal interfaces with annotated regions. The first contrasts an ACL table without errors with a design in which a red command failure conflicts with the displayed ACL state. The second contrasts a rename field without editing instructions with one that exposes contextual editing controls. The third contrasts an appended line shown only within card details with a design that explicitly confirms which file was updated.}
    \label{fig:paired-tui-cases}
\end{figure*}

\section{Discussion}
\label{sec:discussion}

\subsection{Case Study: How TUI Design Shapes Interaction Cost}
\label{sec:generated-tui-case-study}
To illustrate more concretely how TUI design shapes interaction cost, Fig.~\ref{fig:paired-tui-cases} presents three representative paired cases from actual task executions. They cover information visibility, action discoverability, and outcome feedback, matching those in Fig.~\ref{fig:intro_motivation}: blind scanning versus guided navigation, hidden affordances versus visible prompts, and silent failures versus immediate feedback.

\textbf{Information visibility} concerns how clearly the interface presents task-relevant content and system status. P10/T07 asks for Redis ACL users and their rules. Both implementations complete all three runs, but one presents the requested accounts in a table without conflicting messages. The other displays the ACL state while a red \texttt{unknown command} message reports that an ACL operation failed. Mean interaction demand rises from 17.0 to 67.0 AKE and from 3.0k to 21.7k reasoning tokens. The visible failure makes otherwise readable information harder to trust and interpret.

\textbf{Action discoverability} concerns how readily the interface reveals the controls needed for a task. In P03/T04, one design opens a rename field without revealing how to replace its contents. The other exposes clearing, segment deletion, cursor movement, confirmation, and cancellation in the active mode. Both implementations complete all runs, yet the path with contextual guidance falls from 35.0 to 16.0 AKE and from 15.0k to 0.5k reasoning tokens. The gain comes from exposing the editing method at the point of action rather than from adding functionality.

\textbf{Outcome feedback} concerns how clearly the interface communicates the result of an action. P02/T04 asks the user to append a line to the \texttt{FLOW-2} card. Both implementations complete all runs, but one exposes the added text only in the open card. The other also displays the updated card and a checkmarked message identifying \texttt{FLOW-2.md} as the modified file. With this confirmation, mean interaction demand is 14.0 rather than 45.7 AKE and 1.5k rather than 61.4k reasoning tokens. The update can be verified without inferring persistence from the displayed content.

These cases show how interaction cost emerges from concrete interface design choices. Conflicting status information makes visible content harder to trust, contextual guidance reduces trial and error, and explicit confirmation removes ambiguity about whether an update persisted. Weakness in any of these dimensions can increase interpretation or operation cost even when task completion is unchanged. This mechanism-level view explains why the three measures separate in RQ3~\cite{DBLP:conf/chi/Nielsen94b,DBLP:books/daglib/0014798}.

\subsection{Implications}
\subsubsection{For Researchers}
Our findings suggest three research directions:

\textbf{Evaluate usability through task execution.} Functional correctness shows whether a required state can be reached, but not how the interface supports users in reaching it. This study begins with TUIs because they combine GUI-like visual states with CLI-like keyboard interaction that can be recorded and replayed. The same approach can be extended to web, mobile, desktop, and other interactive software when observations, actions, and outcomes can be captured and verified. This study provides an initial human calibration. An important next step is to test whether the observed relationships transfer across interface types, tasks, users, and evaluators.

\textbf{Broaden the evidence captured by automated evaluation.} \AgentAsUser currently provides execution-grounded evidence of task completion and interaction demand. It does not directly capture subjective experiences such as satisfaction and confusion, learning over time, or accessibility for users with different abilities. An important direction is to develop automated evidence for these aspects and determine where human evaluation remains necessary.

\textbf{Diagnose usability problems, not only score interfaces.} A promising research direction is to develop methods that trace increases in interpretation or operation demand to specific interface states and interaction steps. These methods could classify causes such as poor visibility, discoverability, or feedback and return actionable evidence to developers or coding agents. Such diagnosis could support iterative generation and revision while preserving task completion.

\subsubsection{For Developers}
Our findings suggest two development practices:

\textbf{Make interaction quality an explicit acceptance criterion.} Implementations with the same functional scope can differ substantially in completion and interaction demand. During requirements and acceptance testing, development teams can pair functional tests with representative tasks that specify what users must see, which actions must be discoverable, and how outcomes are confirmed. They can then evaluate completion, interpretation, operation, and trajectory evidence together rather than accept an implementation based on its final state alone.

\textbf{Use execution-grounded evaluation to prevent usability regressions.} All visual interfaces require usability evaluation. \AgentAsUser provides quantitative task-execution evidence that can complement human evaluation for a more comprehensive assessment of usability. Once representative tasks, outcome checks, and baseline costs exist, development teams can rerun them to detect interaction regressions even when functional tests continue to pass. The same comparisons can rank candidate implementations in generated-software workflows. Together, automated checks and human evaluation can inform development and release decisions.

\subsection{Limitations and Future Work}
\noindent\hspace*{\parindent}\textbf{Human calibration.} Our human study recruits six experienced terminal users and cannot cover all user populations. Nielsen and Landauer's model suggests diminishing returns after the first few users in homogeneous formative usability studies~\cite{DBLP:conf/chi/NielsenL93}. Our six participants therefore provide an initial homogeneous calibration rather than population-wide validation. The main conclusions are further supported by hierarchical bootstrap inference with 95\% confidence intervals from 10,000 resamples over repositories, tasks, participants, and agent runs, and the cost relationships recur across all five evaluator configurations. Validation with more diverse users and counterbalanced task orders remains future work.

\textbf{Measurement scope.} 
Usability spans effectiveness, efficiency, and satisfaction for specified users, goals, and contexts~\cite{iso9241112018}, and no single measure captures it fully~\cite{DBLP:journals/ijmms/Hornbaek06}. We focus on task-based usability, using task completion and interaction demand as reproducible execution evidence. This scope excludes subjective experience, learning, and accessibility. The evaluator experiments test a common paradigm rather than rank models.
The \KeyboardAgent is intentionally minimal, with fresh context and no external tools, skills, or complex context management. This design keeps the agent as closely aligned with a human user as possible and reduces the influence of other agent configurations. We also repeat each task three times and report five configurations separately. Objective measurement of additional usability dimensions remains future work.

\textbf{Benchmark scope.} The 84 tasks from 15 repositories cannot cover all TUI domains and workflows. This size follows from the screening, manual task design, and oracle validation in Section~\ref{sec:benchmark-construction}, which prioritize reproducible execution, reliable oracles, and recent repositories with less risk of pretraining overlap. Because task mining also had to remain reproducible, these tasks cannot cover every function of the original TUIs, which is unavoidable. Future work can utilize task-mining methods to expand the dataset. Moreover, this study is a first step toward execution-grounded measurement of task-based TUI usability. It can be applied to other interactive settings such as web and mobile applications.

\section{Related Work}
\label{sec:related}

\subsection{Automated and Agent-Based Usability Evaluation}
KLM and GOMS predict skilled human performance by decomposing a specified method into elementary operators~\cite{card1980keystroke,card2018psychology,DBLP:journals/tochi/JohnK96a}. CogTool partially automates this analysis but still produces predicted execution time~\cite{DBLP:conf/chi/JohnPSK04}. These models assume a largely error-free expert path and therefore do not capture exploration or recovery on a newly encountered interface. Heuristic inspection and earlier automation likewise complement participant studies rather than replace them~\cite{DBLP:conf/chi/Nielsen94b,DBLP:journals/csur/IvoryH01}.

Recent LLM systems mainly simulate users, collect think-aloud reports, predict ratings, or score interfaces as judges~\cite{DBLP:conf/chi/0003YGHWLG0LW25,DBLP:conf/hicss/GriesbachSWG26,DBLP:journals/corr/abs-2601-15777,DBLP:journals/corr/abs-2606-05697,DBLP:journals/corr/abs-2604-26020,liu2026iwc}. Questionnaires and preference studies similarly capture subjective experience rather than verified task execution. These approaches provide useful evidence, but they do not measure interaction cost from verified task executions. \AgentAsUser instead proposes an execution-grounded paradigm in which a fixed evaluator executes predefined tasks, deterministic oracles verify completion, and \AgentKLM records interpretation and operation costs along the realized trajectory rather than a judged usability score.

\subsection{Evaluating Agents and Generated Interactive Software}
Computer-use and terminal benchmarks such as OSWorld, Terminal-Bench, and TUA-Bench evaluate whether agents can complete tasks in real environments~\cite{DBLP:conf/nips/XieZCLZCHCSLLXZ24,DBLP:conf/iclr/MerrillSCLRBSSW26,DBLP:journals/corr/abs-2606-28480}. Their target construct is agent capability, and programmatic oracles separate traces from completion judgments~\cite{DBLP:journals/tse/BarrHMSY15}. Recent TUI testing work compares LLM-guided and random exploration for coverage and crash discovery~\cite{DBLP:journals/corr/abs-2608-03743}. Our oracles instead encode user goals, and \AgentKLM characterizes the interaction cost of reaching those goals. Generated-interface work such as AUI-Gym uses computer-use agents to judge newly generated GUIs by task solvability and navigation success, then redesigns those sites toward agent-native completion~\cite{DBLP:journals/corr/abs-2511-15567}. Agent success on a generated website is a design objective, not a measurement of TUI interaction demand.
Prior work therefore shows that agents can operate interfaces, generate UX feedback, and test interactive software. The remaining gap is a repeatable way to measure TUI interaction cost from verified task executions without an LLM usability score. \FrameworkName addresses this gap through actor-independent oracles, atomic keyboard traces, agent-native interpretation expenditure, human relationship characterization, and observation-representation sensitivity.
\section{Conclusion}
In this paper, we conduct an empirical study of \AgentAsUser for measuring task-based TUI usability. We introduce \AgentKLM to separate interpretation and operation costs, and develop \FrameworkName and \BenchmarkName to support reproducible evaluation. We evaluate them across 84 tasks, 15 reference TUIs, six human participants, five observation--evaluator configurations, and 45 generated counterparts. We find weak human--agent completion correspondence but aligned cost rankings, especially for operation. This pattern persists across observation settings. Generated TUIs can complete the required tasks, yet their task-based usability still needs improvement. These findings position Agent-as-a-User as a repeatable and execution-grounded paradigm for measuring task-based usability.

\section*{Data Availability}
All artifacts for \FrameworkName and \BenchmarkName, including the framework implementation and benchmark materials, are available at \url{https://github.com/kinesiatricssxilm14/Agent-as-a-User}. To facilitate inspection, we additionally provide anonymized trajectories for all runs at \url{https://tuinaut.pages.dev/}.


\bibliographystyle{ACM-Reference-Format}
\bibliography{references}

\end{document}